\documentclass[a4paper]{article}

\usepackage{hyperref}
\usepackage{dsfont}
\usepackage{color}
\usepackage{geometry}

\usepackage{amsmath,amsthm,amssymb,epsfig,graphicx}
\usepackage[latin1]{inputenc}

\usepackage{xspace}
\usepackage{latexsym}
\usepackage{amsmath,amstext,amsxtra,amsgen,amsbsy,amsopn,amscd}
\usepackage{latexsym}
\usepackage{mathrsfs}
\usepackage{dsfont}
\usepackage[active]{srcltx}

\numberwithin{equation}{section}

\newcommand{\ii}{\infty}
\newcommand{\Z}{\mathbb{Z}}
\newcommand{\R}{\mathbb{R}}
\newcommand{\N}{\mathbb{N}}

\newcommand{\C}{\mathbb{C}}

\newcommand{\E}{\mathcal{E}}
\newcommand{\cE}{\mathcal{E}}

\newcommand{\cF}{\mathcal{F}}
\newcommand{\cL}{\mathscr{L}}

\newcommand{\Q}{\mathcal{Q}}

\newcommand{\cP}{\mathcal{P}}
\newcommand{\cB}{\mathcal{B}}

\newcommand{\gS}{\mathfrak{S}}
\newcommand{\gH}{\mathfrak{H}}

\newcommand{\norm}[1]{ \left| \! \left| #1 \right| \! \right| }
\newcommand{\wto}{\rightharpoonup}
\renewcommand{\epsilon}{\varepsilon}

\def\XXint#1#2#3{{\setbox0=\hbox{$#1{#2#3}{\int}$}
     \vcenter{\hbox{$#2#3$}}\kern-.5\wd0}}

\DeclareMathOperator{\Tr}{{\rm Tr}}
\DeclareMathOperator{\Tro}{{\rm Tr}_0}
\newcommand{\tr}{\Tr}
\newcommand{\bral}{\left<}
\newcommand{\brar}{\right|}
\newcommand{\ketl}{\left|}
\newcommand{\ketr}{\right>}

\newcommand{\dem}{\varepsilon_{\rm M}}

\newcommand{\ptgf}{\mathcal{E}^{\rm P}_{\dem}}
\newcommand{\ptge}{E^{\rm P}_{\dem}}
\newcommand{\ptgm}{\psi^{\rm P}_{\dem}}

\newcommand{\ptgint}{F^{\rm P}_{\dem}}

\newcommand{\FSm}{\gamma_{\rm per} ^0}

\newcommand{\FSh}{H_{\rm per} ^0}
\newcommand{\FSl}{\epsilon_{\rm F}}
\newcommand{\FSd}{\rho_{\rm per} ^0}

\newcommand{\FSp}{V_{\rm per} ^0}

\newcommand{\Wrho}{W_{\rho}}

\newcommand{\rhoQ}{\rho_Q}

\newcommand{\rhoP}{\rho_{\Psi}}

\newcommand{\tote}{E_m}

\newcommand{\crysf}{\mathcal{F}_{\rm crys} }
\newcommand{\cryse}{F_{\rm crys} }

\newcommand{\Eperf}{\mathcal{E}^{\rm per}_m}
\newcommand{\Epere}{E ^{\rm per}_m}
\newcommand{\Eperm}{u_m^{\rm per}}

\newcommand{\Eperelim}{E ^{\rm per}}

\newcommand{\Qpp}{Q ^{++}}
\newcommand{\Qpm}{Q ^{+-}}
\newcommand{\Qmp}{Q^{-+}}
\newcommand{\Qmm}{Q^{--}}

\newcommand{\one}{{\ensuremath {\mathds {1}} }}
\newcommand{\oneep}{\one_{\left(-\infty, \FSl \right)}}

\newcommand{\coulker}{|\: . \:| ^{-1}}
\newcommand{\rhogam}{\rho_{\gamma}}

\newcommand{\munucper}{\mu ^{0} _{\rm per}}

\newcommand{\rhff}{\E ^{\rm rHF}}

\newcommand{\rhfe}{E ^{\rm rHF}}

\newcommand{\Sch}{\mathfrak{S}}

\newcommand{\Ex}{\mathrm{Ex}}

 \newtheorem{theorem}{Th\'eor\`eme}[section]
 \newtheorem{proposition}{Proposition}[section]

\numberwithin{equation}{section}

\title{Sur la mod\'elisation de l'interaction entre polarons et cristaux quantiques\footnote{S\'eminaire Laurent Schwartz ``EDP et applications'', donn\'e le 18 D\'ecembre 2012 \`a l'Ecole Polytechnique.}}
\author{N. Rougerie \\
\normalsize \it Universit\'e Grenoble 1 et CNRS, LPMMC, UMR 5493, BP 166, 38042 Grenoble, France.\\
}

\date{$1^{\rm er}$ Juin 2013}

\begin{document}

\maketitle

\begin{abstract}
Je r\'esume dans ce texte des travaux r\'ecents, en collaboration avec Mathieu Lewin, sur la mod\'elisation des (multi-)polarons. Il s'agit de d\'ecrire le syst\`eme physique form\'e par l'interaction entre une ou plusieurs particules charg\'ees et un cristal constitu\'e d'un nombre infini de noyaux classiques et d'\'electrons quantiques. Nous d\'efinissons un nouveau mod\`ele  en couplant l'\'equation de Schr\"odinger pour les particules charg\'ees avec un mod\`ele de type Hartree-Fock r\'eduit d\'ecrivant la r\'eaction des \'electrons du cristal. Nous \'etudions l'existence d'\'etats li\'es (minimiseurs de la fonctionnelle d'\'energie) et d\'emontrons que le mod\`ele de Pekar pour le grand polaron peut se d\'eduire du n\^otre dans une limite macroscopique. \\

\end{abstract}

\tableofcontents

\bigskip

En physique de la mati\`ere condens\'ee on appelle \emph{polaron} le syst\`eme form\'e par l'interaction entre un cristal quantique et une particule charg\'ee (en g\'en\'eral un \'electron) (voir \cite{AleDev-09} pour des r\'ef\'erences dans la litt\'erature physique). Similairement, on appelle \emph{multi-polaron} le syst\`eme form\'e par l'interaction entre un cristal quantique et plusieurs particules charg\'ees. Un fait remarquable est la possibilit\'e d'obtenir des \'etats li\'es dans un tel syst\`eme. En effet, dans le vide, les \'electrons (ou n'importe quel type de particules de m\^eme charges) se repoussent et la formation d'\'etats li\'es n'est donc jamais favorable \'energ\'etiquement. L'interaction avec le cristal r\'esulte en une force attractive qui permet aux \'electrons de s'agglom\'erer en un \'etat stable. Le m\'ecanisme est grossi\`erement le suivant: un \'electron cr\'ee une polarisation locale des atomes du cristal en attirant leurs nucl\'eons et repoussant leurs \'electrons. Cette polarisation locale autour d'un \'electron sera \`a son tour ressentie par les autres \'electrons, qui seront ainsi en position d'interagir entre eux via le cristal. L' interaction effective r\'esultant de ce m\'ecanisme est attractive et peut dans certains cas contrebalancer la r\'epulsion (de type Coulomb) existant naturellement entre les \'electrons.

Le polaron est la particule effective form\'ee par l'\'electron ajout\'e au cristal et le champ de polarisation qu'il y cr\'ee. On parle parfois d'\'electrons habill\'es (par le champ de polarisation), avec de nouvelles propri\'et\'es physiques effectives. Dans cet expos\'e je ferais un abus de langage en appelant ``polarons'' les \'electrons ajout\'es au cristal. Cela permettra de les distinguer des \'electrons des atomes du cristal qui ne seront pas trait\'es de la m\^eme mani\`ere, comme discut\'e plus bas. 

Cet expos\'e est consacr\'e \`a la mod\'elisation pr\'ecise de l'interaction effective polaron-cristal. Je r\'esume les travaux \cite{LewRou-11,LewRou-12} o\`u nous avons introduit un nouveau mod\`ele de polaron en couplant l'\'equation de Schr\"odinger \`a $N$ corps avec un mod\`ele de cristal quantique avec d\'efaut introduit dans \cite{CanDelLew-08a,CanDelLew-08b,CanLew-10} (dans la lign\'ee du travail \cite{CatBriLio-01} sur le cristal parfait). Nous d\'emontrons tout d'abord l'existence d'\'etats li\'es (minimiseurs de la fonctionnelle d'\'energie) pour le cas d'un polaron, et un th\'eor\`eme de type HVZ pour le cas du multi-polaron, c'est-\`a-dire l'existence d'\'etats li\'es en supposant la validit\'e d'in\'egalit\'es de liaisons quantifi\'ees. Dans un deuxi\`eme temps nous montrons que le fameux mod\`ele de Pekar pour le grand polaron peut se d\'eduire du n\^otre dans une limite macroscopique (taille caract\'eristique du cristal tendant vers l'infini ou, de mani\`ere \'equivalente, masse du polaron tendant vers z\'ero).

\medskip

Dans une premi\`ere section, j'explique plus en d\'etail le concept de polaron, donne une forme g\'en\'erique possible pour la fonctionnelle d'\'energie et discute le cas le plus simple rentrant dans ce cadre, c'est-\`a-dire le mod\`ele de Pekar. Une deuxi\`eme section d\'ecrit le mod\`ele de cristal quantique que nous utiliserons, r\'esumant les travaux~\cite{CatBriLio-01,CanDelLew-08a,CanDelLew-08b,CanLew-10}. Notre mod\`ele de polaron est pr\'esent\'e dans la Section 3, avec les r\'esultats d'existence d'\'etats li\'es associ\'es, prouv\'es dans \cite{LewRou-12}. Enfin, la Section 4 r\'esume le travail \cite{LewRou-11} et explique comment retrouver le mod\`ele de Pekar dans une limite macroscopique.

\bigskip

\emph{Les travaux r\'esum\'es dans cet expos\'e ont \'et\'e financ\'es par l'European Research Council : European Community's Seventh Framework Programme (FP7/2007-2013 Grant Agreement MNIQS No. 258023).}

\section{Le concept de polaron}

G\'en\'eriquement, on obtient un mod\`ele de polaron en couplant un mod\`ele pour les \'electrons dans le vide avec un mod\`ele d\'ecrivant les excitations d'un cristal. Le Hamiltonien pour $N$ \'electrons tri-dimensionnels non relativistes dans le vide est (en version adimensionalis\'ee)
\begin{equation}\label{eq:schro N corps}
H_N :=\sum_{j=1} ^N -\frac{1}{2} \Delta_{x_j} + \sum_{1\leq i < j \leq N} \frac{1}{|x_i-x_j|}
\end{equation}
agissant sur $\bigotimes_{j=1} ^N L ^2 (\R ^3)$, le produit tensoriel de $N$ copies de l'espace de Hilbert pour une particule, $L ^2 (\R ^3)$. Comme d'habitude on identifiera cet espace avec $L ^2 (\R ^{3N})$.  Pour des \'electrons, qui sont des particules fermioniques, il faut de plus imposer une contrainte d'antisym\'etrie (principe d'exclusion de Pauli) sur les fonctions d'onde $\Psi$ d\'ecrivant les $N$ particules:
\begin{equation}\label{eq:antisym}
 \Psi(x_1,\ldots,x_N) = \mathrm{sgn} (\sigma) \Psi (x_{\sigma (1)}, \ldots, x_{\sigma (N)})
\end{equation}
pour toute permutation $\sigma$ des $N$ particules, avec $\mathrm{sgn} (\sigma)$ la signature de la permutation. En d'autres termes, on consid\`ere l'action  du Hamiltonien sur le produit tensoriel antisym\'etrique de $N$ copies de $L ^2 (\R ^3)$. Pour les r\'esultats que nous pr\'esentons, le principe de Pauli ne joue en fait pas un r\^ole essentiel et on pourra ignorer la contrainte \eqref{eq:antisym}, ou m\^eme la remplacer par une contrainte de sym\'etrie (rempla\c{c}ant $\mathrm{sgn}  (\sigma)$ par $1$ dans \eqref{eq:antisym}) ce qui revient \`a consid\'erer des particules bosoniques plut\^ot que fermioniques. 

La fonctionnelle d'\'energie pour des \'electrons dans le vide, d\'ecrits par une fonction d'onde \`a $N$ corps $\Psi \in L ^2 (\R ^{3N})$ est donn\'ee par 
\begin{equation}\label{eq:ener vide}
\cE [\Psi]:=  \bral \Psi, H_N \Psi \ketr_{L ^2}.
\end{equation}
Comme toujours, $|\Psi| ^2$ est interpr\'etée comme une densit\'e de probabilit\'e de trouver le syst\`eme dans un certain \'etat. Les fonctions d'onde admissibles satisfont donc 
\begin{equation}\label{eq:contrainte masse}
\left\Vert \Psi \right\Vert_{L ^2 (\R ^{3N})} = 1 
\end{equation}
et au vu du caract\`ere r\'epulsif des interactions (deuxi\`eme terme dans \eqref{eq:schro N corps}) la fonctionnelle \eqref{eq:ener vide} n'a pas de minimiseur sous cette contrainte, ce qui revient \`a dire que les \'electrons ne forment jamais d'\'etats li\'es dans le vide.

Comme on le voit, aucun effort de mod\'elisation particulier n'est fait pour les \'electrons que l'on rajoute au cristal: on utilise le mod\`ele le plus fondamental, et a priori le plus complexe imaginable pour d\'ecrire l'interaction d'\'electrons non relativistes. 

\medskip

La description du cristal est plus d\'elicate. On pense g\'en\'eralement \`a un cristal tr\`es \'etendu dans lequel les polarons \'evoluent. Pour \'eviter des effets de bord, on imagine donc un cristal infini, et il s'agit alors de d\'ecrire un syst\`eme compos\'e d'un nombre infini de particules. Il faut ensuite d\'ecrire la r\'eaction de ce syst\`eme \`a l'insertion de une ou plusieurs particules charg\'ees suppl\'ementaires (les polarons), et enfin coupler cette description avec le Hamiltonien \eqref{eq:schro N corps}. 

On est souvent amen\'e \`a utiliser une description tr\`es simplifi\'ee du cristal, en particulier lorsqu'il s'agit de d\'emontrer des r\'esultats math\'ematiques rigoureux. La simplification la plus extr\^eme consiste \`a voir le cristal comme un milieu di\'electrique classique continu, ce qui conduit au mod\`ele de Pekar d\'ecrit plus bas.

\medskip

Nous adopterons dans cet expos\'e les hypoth\`eses de mod\'elisation suivantes (et en ferons quelques autres au fur et \`a mesure):

\medskip

\noindent(H1)\emph{ Absence de corr\'elations entre le ou les polarons et le milieu di\'electrique ambiant.}\\
(H2) \emph{Les \'electrons en surplus qui constituent le multi-polaron sont distinguables des \'electrons du cristal.}\\
(H3) \emph{Les noyaux du cristal sont trait\'es classiquement.}

\medskip

La deuxi\`eme hypoth\`ese est tr\`es difficile \`a contourner, puisqu' implicite dans le concept de polaron tel que nous l'avons d\'ecrit informellement plus haut. La troisi\`eme est parfaitement naturelle au vu de la grande disproportion entre la masse des noyaux et celle des \'electrons (approximation de Born-Oppenheimer). La premi\`ere hypoth\`ese est d'une nature plus subtile et est certainement discutable du point de vue physique. On peut la voir comme une hypoth\`ese de couplage fort entre les polarons et le cristal. Dans le cadre du mod\`ele de Fr\"ohlich (auquel nous ferons allusion plus bas), on peut d\'emontrer rigoureusement cette connexion entre couplage fort et absence de corr\'elations \cite{DonVar-83,LieTho-97}.

\subsection{Forme g\'en\'erique possible d'un mod\`ele de polaron}\label{sec:generic}

Dans cette section nous d\'ecrivons informellement le genre de fonctionnelle qu'il nous faudra construire et \'etudier dans la suite. On suppose un cristal occupant tout l'espace pour ne pas avoir \`a se soucier de conditions de bord. Il est constitu\'e de noyaux classiques d\'ecrits par d'une distribution de charge $\cL$-p\'eriodique $\mu^0_{\rm per}\geq0$ (avec $\cL$ un r\'eseau de $\R ^3$) et d'\'electrons quantiques d\'ecrits par une densit\'e $\cL$-p\'eriodique $\rho^0_{\rm per}$. On notera $\Gamma$ la cellule unit\'e du r\'eseau, un exemple simple \'etant simplement $\cL=\Z ^3$, avec $\Gamma$ le cube unit\'e. Au repos le cristal est localement neutre :
\begin{equation}\label{eq:neutral local}
\int_{\Gamma}\rho^0_{\rm per}=\int_{\Gamma}\mu^0_{\rm per} 
\end{equation}
et produit un potentiel \'electrostatique $V^0_{\rm per}$:
$$-\Delta V^0_{\rm per}=4\pi\big(\rho^0_{\rm per}-\mu^0_{\rm per}\big)$$
que ressentira toute nouvelle particule ins\'er\'ee dans le cristal. Par la suite on sera amen\'e \`a d\'ecrire les \'electrons plus finement, par une matrice de densit\'e plut\^ot qu'une simple densit\'e de charge comme ici. Pour simplifier l'expos\'e nous commen\c{c}ons par ce cas plus simple, auquel il y aura toujours moyen de se ramener, quoique de mani\`ere un peu artificielle (voir Section \ref{sec:model}). 

Quand une nouvelle particule charg\'ee est ajout\'ee au syst\`eme, elle d\'eplace les noyaux et les \'electrons du cristal. Pour des raisons d'ad\'equation entre la technologie math\'ematique actuelle et le ph\'enom\`ene physique en question nous ferons une nouvelle hypoth\`ese de mod\'elisation:

\medskip

(H4) \emph{Les noyaux du cristal ne sont pas d\'eplac\'es par les polarons.}

\medskip

Il s'agit d'une hypoth\`ese discutable mais la relaxer am\`ene des difficult\'es math\'ematiques importantes: il faudrait pouvoir démontrer que la distribution p\'eriodique des noyaux est bien optimale \'energ\'etiquement, de sorte que les perturbations autour ce cet arrangement soient petites et quantifiables. Étant donn\'e que ceci est un des plus fameux probl\`emes ouverts en math\'ematiques et en physique (probl\`eme de cristallisation), nous ne nous aventurerons pas de ce c\^ot\'e l\`a et ferons toujours l'hypoth\`ese $4$. Ceci implique que nous n'obtiendrons dans nos mod\`eles que la contribution des \'electrons aux propri\'et\'es di\'electriques du cristal, mais pour ce qui concerne par exemple le mod\`ele de Pekar, cela n'influera pas la forme de la fonctionnelle obtenue, seulement les valeurs de ses param\`etres.  

Nous d\'ecrivons donc l'effet des polarons sur le cristal par une perturbation de la densit\'e des \'electrons:
$$ \FSd \to \rho=\rho^0_{\rm per}+\delta\rho.$$ 
Les polarons ressentiront la perturbation du potentiel \'electrique \footnote{$\star$ repr\'esente la convolution} 
$$\delta\rho \, \star \, |\, . \,|^{-1}$$
induite par le d\'eplacement des \'electrons, et c'est cette r\'etroaction qui induit la force attractive effective entre les polarons. Bien s\^ur, d\'eplacer les \'electrons du cristal de leurs positions d'\'equilibre a un certain co\^ut \'energ\'etique que nous noterons $\cF_{\rm crys}[\delta\rho]$. Tout l'effort de mod\'elisation portera sur la d\'efinition d'une fonctionnelle appropri\'ee, qui soit raisonnable \`a la fois  du point de vue du r\'ealisme physique et de celui de la rigueur math\'ematique, ce qui demandera des compromis bien choisis. 

Pour un seul polaron d\'ecrit par une fonction d'onde $\psi \in L ^2 (\R ^3)$ nous arrivons donc \`a la forme suivante pour l'\'energie du syst\`eme perturb\'e:
\begin{multline}
\cE[\psi,\delta\rho]=\frac1{2}\int_{\R^3}|\nabla\psi(x)|^2\,dx+\int_{\R^3}V^0_{\rm per}(x)|\psi(x)|^2\,dx\\+\int_{\R^3}\int_{\R^3}\frac{\delta\rho(x)|\psi(y)|^2}{|x-y|}\,dx\,dy+\cF_{\rm crys}[\delta\rho].
\label{eq:abstract-energy}
\end{multline}
Le premier terme est l'énergie cin\'etique du polaron dans le vide, le deuxi\`eme l'interaction entre le polaron et le champ \'electrique du cristal au repos, le troisi\`eme l'interaction Coulombienne (classique) entre le polaron et la perturbation $\delta \rho$ des \'electrons du cristal, le quatri\`eme le co\^ut \'energ\'etique pour cr\'eer la perturbation. 

En \'ecrivant une fonctionnelle de deux variables s\'epar\'ees $\psi$ et $\delta \rho$ nous avons suivi notre hypoth\`ese $1$ de d\'ecorr\'elation entre le comportement du polaron et celui du cristal. Dans ce cadre il est possible de compl\`etement \'eliminer les degr\'es de libert\'e du cristal en minimisant par rapport \`a $\delta \rho$ pour un $\psi$ fix\'e. On obtient ainsi une fonctionnelle effective pour le seul polaron
\begin{equation}
\cE_{\rm eff}[\psi]=\frac1{2}\int_{\R^3}|\nabla\psi(x)|^2\,dx+\int_{\R^3}V^0_{\rm per}(x)|\psi(x)|^2\,dx+F_{\rm crys}\big[|\psi|^2\big],
\label{eq:Pekar-abstract-intro}
\end{equation}
avec 
\begin{equation}
 F_{\rm crys}\big[|\psi|^2\big]=\inf_{\substack{\delta\rho\geq-\rho^0_{\rm per}}}\left(\int_{\R^3}\int_{\R^3}\frac{\delta\rho(x)|\psi(y)|^2}{|x-y|}\,dx\,dy+\cF_{\rm crys}[\delta\rho]\right)
\label{eq:def_nonlinearity_intro}
\end{equation}
Notons la contrainte $\delta \rho \geq - \rho^0_{\rm per}$ qui exprime simplement le fait que la densit\'e de charge perturb\'ee des \'electrons $ \rho^0_{\rm per} + \delta \rho$ doit garder le bon signe puisqu'on ne fait que d\'eplacer des charges. 

Dans le cas d'un $N$-polaron d\'ecrit par une fonction d'onde \`a $N$-corps  $\Psi \in L ^2 (\R ^{3N})$ il faut aussi tenir compte des interactions Coulombiennes entre chaque paire d'\'electrons, d\'ecrites par le Hamiltonien \eqref{eq:schro N corps}: 
\begin{multline}
\cE_{\rm eff}[\Psi]=\int_{\R^{3N}}\left(\frac1{2}\sum_{j=1}^N\left|\nabla_{x_j}\Psi(x_1,...,x_N)\right|^2+\sum_{1\leq k<\ell\leq N}\frac{|\Psi(x_1,...,x_N)|^2}{|x_k-x_\ell|}\right)dx_1\cdots dx_N\\
+\int_{\R^3}V^0_{\rm per}(x)\rho_{\Psi}(x)\,dx+F_{\rm crys}\big[\rho_{\Psi}\big]
\label{eq:Pekar-N-abstract-intro}
\end{multline}
o\`u 
\begin{equation}\label{eq:rhoP}
 \rhoP (x):= \int_{\R ^{3(N-1)}} |\Psi (x,x_2\ldots,x_N)| ^2 dx_2 \ldots dx_N  
\end{equation}
est la densit\'e de charge associ\'ee au $N$-polaron (vue la contrainte \eqref{eq:antisym}, le choix des variables sur lesquelles on int\`egre $|\Psi| ^2$ n'a pas d'importance). Nous tenons compte des corr\'elations possibles entre les polarons puisque nous les d\'ecrivons par une fonction d'onde \`a $N$ corps plut\^ot que par une collection de $N$ fonctions \`a $1$ corps. En fait il se trouve que les \'etats li\'es entre polarons, lorsqu'ils existent, sont purement dus aux corr\'elations, par exemple sous forme de forces de Van der Waals, voir \cite[Section 5.3]{Lewin-11}. Il n'y a donc pas de simplification raisonnable \`a faire \`a ce niveau.

\medskip

La t\^ache la plus difficile du programme esquiss\'e ci-dessus est la d\'efinition de la fonctionnelle $F_{\rm crys}\big[\rho\big]$ d\'ecrivant la r\'eponse du cristal \`a l'insertion d'une nouvelle densit\'e de charge $\rho$. Cette fonctionnelle devra avoir un caract\`ere math\'ematiquement bien pos\'e au sens de l'existence d'un infimum dans \eqref{eq:def_nonlinearity_intro}, et si possible d'un minimiseur.  Le paragraphe suivant est consacr\'e \`a la fonctionnelle sans doutes la plus simple (mais d\'ej\`a largement utilis\'ee par les physiciens) rentrant dans ce cadre, la fonctionnelle de Pekar.

\subsection{Le mod\`ele de Pekar pour le grand polaron}

Dans le cadre du mod\`ele de Pekar, on suppose que le cristal peut \^etre approxim\'e par un milieu di\'electrique classique, continu et isotrope, avec constante di\'electrique $\dem>1$.  Vu le caract\`ere discret d'un cristal v\'eritable il est clair qu'une telle description n'a de chances d'\^etre valable que dans le cas d'un  \emph{grand polaron}, c'est-\`a-dire d'un polaron vivant sur une \'echelle de longueur tr\`es grande devant la cellule unit\'e du cristal sous-jacent.  Dans ce cas, la contrainte de neutralit\'e locale \eqref{eq:neutral local} sugg\`ere que l'influence du potentiel \'electrique du cristal au repos $\FSp$ sera n\'egligeable, et nous l'ignorerons donc dans ce paragraphe. Dans cette description simplifi\'ee, le cristal au repos appara\^it (du point de vue du polaron) comme compl\`etement neutre. D\`es lors le polaron ressentira exclusivement l'influence de la perturbation $\delta \rho$ qu'il induit dans le milieu apparemment neutre dans lequel il est ins\'er\'e.

Dans un milieu classique continu le co\^ut pour cr\'eer la perturbation $\delta \rho$ est simplement donn\'e par son \'energie \'electrostatique:
\[
\crysf[\delta \rho] = \frac{\alpha}{2} \int_{\R^3}\int_{\R^3}\frac{\delta \rho (x) \delta \rho(y)}{|x-y|}\,dx\,dy
\]
avec $\alpha = 1- \dem ^{-1}>0$. La minimisation par rapport \`a $\delta \rho$ dans \eqref{eq:def_nonlinearity_intro} est alors explicite et la fonctionnelle \eqref{eq:Pekar-abstract-intro} devient  
\begin{align}\label{eq:Pekar 1}
 \ptgf[\psi]&=  \int_{\R ^{3}} \frac{1}{2}|\nabla \psi| ^2 dx  - \frac{\alpha}{2} \iint_{\R^3 \times \R ^3} \frac{\left|\psi (x) \right|^2 \left| \psi  (y)\right|^2 }{|x-y|}dx dy \nonumber\\
&= \int_{\R ^{3}} \frac{1}{2}|\nabla \psi| ^2 dx  + \ptgint[\rho]
\end{align}
qui se g\'en\'eralise en  
\begin{equation}
 \label{eq:Pekar N}
 \ptgf [\Psi] = \sum_{i=1} ^N \int_{\R ^{3N}} \frac{1}{2} |\nabla_i \Psi| ^2 dX + \sum_{i<j} \int_{\R^{3N}} \frac{| \Psi (X)| ^2}{|x_i - x_j|}dX + \ptgint [\rhoP]
\end{equation}
pour le cas du $N$-polaron. Le caract\`ere attractif de l'interaction effective induite par le cristal est \'evidente dans ce cas (de mani\`ere g\'en\'erale $\alpha = 1- \dem ^{-1}>0$ pour tout milieu di\'electrique).

\medskip

Pour des raisons p\'edagogiques, notre introduction du mod\`ele de Pekar est un brin na\"ive. On le consid\`ere souvent~\cite{Pekar-54,Pekar-63,PekTom-51} comme une simplification du fameux mod\`ele de Fr\"ohlich~\cite{Frohlich-37,Frohlich-52} o\`u les excitations du cristal sont mod\'elis\'es par des phonons, c'est-\`a-dire des modes vibratoires du r\'eseau atomique. Dans ce cadre il est usuel de prendre en compte les corr\'elations possibles entre polarons et phonons, et le syst\`eme complet est d\'ecrit par un vecteur de $L ^2 (\R ^{3N}) \otimes \mathcal{F}$ o\`u $\mathcal{F}$ d\'esigne l'espace de Fock (bosonique) pour les phonons. Vu que les phonons d\'ecrivent de petits d\'eplacement des particules du cristal, l'interaction phonons-polarons est alors lin\'eaire et de nature dipolaire, et le co\^ut \'energ\'etique pour cr\'eer un phonon de moment $\mathbf{k}$ est consid\'er\'e comme ind\'ependant de $\mathbf{k}$ (phonons optiques, avec relation de dispersion plate). Dans la limite d'un couplage fort entre polarons et phonons, les corr\'elations disparaissent au sens o\`u un ansatz de la forme $\Psi \otimes \Phi$, avec $\Psi \in L ^2 (\R ^{3N})$ et $\Phi \in \mathcal{F}$, donne une approximation correcte du fondamental du mod\`ele. Le $\Phi$ optimal est alors explicitement donn\'e par un \'etat coh\'erent (voir par exemple \cite{GriMol-10}) et les phonons peuvent \^etre \'eliminés du mod\`ele, ce qui conduit \`a la fonctionnelle de Pekar. Une preuve rigoureuse de cette d\'erivation a \'et\'e donn\'ee d'abord par Donsker et Varadhan \cite{DonVar-83} puis grandement simplifi\'ee par Lieb et Thomas \cite{LieTho-97}. Nous en resterons l\`a pour notre \'evocation du mod\`ele de Fr\"ohlich et renvoyons le lecteur aux travaux \cite{FraLieSeiTho-10,FraLieSeiTho-11,GriMol-10} et leurs r\'ef\'erences pour une pr\'esentation plus compl\`ete et des r\'esultats math\'ematiques.

\medskip

L'existence et l'unicit\'e d'un minimiseur pour la fonctionnelle \eqref{eq:Pekar 1} sont dues \`a Lieb. La pr\'ecompacit\'e des suites minimisantes (et donc l'existence) se d\'eduit \'egalement de la m\'ethode de concentration-compacit\'e de Lions~\cite{Lions-84,Lions-84b}.

\begin{theorem}[\textbf{Minimisation de la fonctionnelle de Pekar, $N=1$, \cite{Lieb-77,Lions-84,Lions-84b}}]\label{theo:PTg1}\mbox{}\\
On suppose que $\dem>1$ 
\begin{enumerate}
\item (\textbf{Existence et unicit\'e}). La fonctionnelle \eqref{eq:Pekar 1} admet un unique minimiseur sous la contrainte de masse $\norm{\psi}_{L ^2} = 1$ qui satisfait 
\begin{equation}\label{equationptg1}
-\Delta \ptgm - \ptgm \left( 1 - \dem ^{-1} \right) |\ptgm| ^2 \star \coulker = \lambda \ptgm
\end{equation}
pour un certain $\lambda \in \R$. 
\item (\textbf{Convergence des suites minimisantes}). Toutes les suite minimisantes $(\psi_n)$ pour \eqref{eq:Pekar 1} sont pr\'ecompactes dans $H ^1 (\R ^3)$, \`a translation pr\`es: il existe une sous-suite $(n_k)_{k\in \N}$, une suite de translations $(\tau_k)\subset\R^3$ et un minimiseur $\ptgm$ tels que 
$$\psi_{n_k}(\cdot-\tau_k)\to\ptgm$$
fortement dans $H^1(\R^3)$.
\end{enumerate}
\end{theorem}

L'\'equation variationnelle \eqref{equationptg1} est parfois appel\'ee \'equation de Schr\"odinger-Newton ou de Choquard. La pr\'esence du multiplicateur de Lagrange $\lambda$ est bien s\^ur due \`a la contrainte de masse sous laquelle s'effectue la minimisation. On voit ici que le mod\`ele de Pekar pr\'edit correctement l'existence des \'etats li\'es d'un polaron: le polaron est pi\'eg\'e dans une perturbation du cristal induite par sa propre densit\'e de charge. 

L'existence d'\'etats li\'es de plusieurs polarons est une question beaucoup plus subtile: pour qu'elle soit possible, l'attraction effective due au terme de Pekar doit contrebalancer la r\'epulsion Coulombienne entre les \'electrons, ce qui n'est possible que pour certaines valeurs des param\`etres $\dem$ et $N$ du mod\`ele. Citons deux r\'esultats, extraits respectivement de \cite{FraLieSeiTho-11} et de \cite{Lewin-11}: 

\begin{theorem}[\textbf{Non-existence de multi-polarons} \cite{FraLieSeiTho-11}]\label{thm:Pekar N non exist}\mbox{}\\
Il existe $\beta > 1$ tel que $\ptgf$ n'a pas de minimiseur pour $\dem < \beta$ et $N \geq 2$.
\end{theorem}

\begin{theorem}[\textbf{Existence de $N$-polarons} \cite{Lewin-11}]\label{thm:Pekar N exist}\mbox{}\\
Pour tout $N \geq 2$, il existe $\beta (N)$ tel que $\ptgf$ a un minimiseur pour $\dem > \beta(N)$.
\end{theorem}

On notera l'absence de d\'ependance en $N$ du $\beta$ donn\'e par le Th\'eor\`eme \ref{thm:Pekar N non exist}, ce qui n'est certainement pas optimal. Combler l'\'ecart entre les deux pr\'ec\'edents r\'esultats constitue un probl\`eme ouvert int\'eressant qui implique l'\'etude d'in\'egalit\'es de liaison (dans l'esprit du principe de concentration-compacit\'e) quantifi\'ees, au vu du r\'esultat suivant: 

\begin{theorem}[\textbf{Existence de multi-polarons et in\'egalit\'es de liaison~\cite{Lewin-11}}]\label{theo:PTgN}\mbox{}\\
On suppose que $\dem>1$. Les propositions suivantes sont \'equivalentes:
\begin{enumerate}
\item (\textbf{In\'egalit\'es de liaisons}). On a  
\begin{equation}\label{binding ptgN}
\ptge (N) < \ptge (N-k) + \ptge (k) \mbox{ pour tout } k=1 \ldots N-1. 
\end{equation}

\smallskip

\item (\textbf{Convergence des suites minimisantes}). Toutes les suites minimisantes pour $\ptge (N)$ sont pr\'ecompactes dans $H ^1 (\R ^{3N})$, \`a translation pr\`es. En particulier, il existe un minimiseur $\Psi^{\rm P}_{\dem}$ pour $\ptge(N)$.
\end{enumerate}
\end{theorem}

Ceci est un r\'esultat de type HVZ (Hunziker-Van Winter-Zhislin~\cite{Hun-66,VanWinter-64,Zhislin-71}), mais pour une fonctionnelle non-lin\'eaire, ce qui est la difficult\'e majeure de sa preuve. Les in\'egalit\'es de liaison \ref{binding ptgN} expriment le fait qu'il n'est pas favorable d'envoyer des particules \`a l'infini, ce qui emp\^eche le cas de dichotomie dans le principe de concentration compacit\'e. 

\section{Mod\`eles de cristaux quantiques}\label{sec:cristal}

Cette section est consacr\'ee \`a la description de la fonctionnelle d'\'energie des \'electrons du cristal. Rappelons que les noyaux sont consid\'er\'es comme fixes, leur distribution est donc une donn\'ee du probl\`eme. Dans un premier temps nous expliquons le cadre g\'en\'eral de la th\'eorie de Hartree-Fock r\'eduite, avant de pr\'esenter le mod\`ele cristallin qui nous int\'eressera plus particuli\`erement, en deux \'etapes: le cristal parfait (p\'eriodique) et le cristal avec d\'efaut. Dans les deux cas, la justification physique des mod\`eles que nous introduisons est bas\'ee sur des arguments de limite thermodynamique \cite{LieSim-77,CatBriLio-98,CatBriLio-01,CanDelLew-08a} : on consid\`ere d'abord un mod\`ele avec un nombre de particules $N$ fini, puis on prend la limite $N\to \infty$ du mod\`ele en fixant le nombre de particules par unit\'e de volume. Les fonctionnelles que nous utiliserons sont celles obtenues \`a la limite, nous renvoyons \`a la litt\'erature pour leur d\'erivation. 
Pour obtenir une fonctionnelle bien d\'efinie, il est n\'ecessaire de simplifier l'\'equation de Schr\"odinger: l'existence de la limite thermodynamique est connue dans le cadre de l'\'equation de Schr\"odinger \cite{LieLeb-72,Fefferman-85,HaiLewSol_1-09,HaiLewSol_2-09}, mais sans expression explicite.  

Comme dans le reste de l'expos\'e nous nous en tiendrons \`a des propri\'et\'es statiques, voir \cite{CanSto-11,CanLewSto-11_proc} pour le cas dynamique.

\subsection{Th\'eorie de Hartree-Fock r\'eduite}

La description du cristal que nous utilisons dans notre mod\`ele de polaron est de type Hartree-Fock r\'eduite: l'\'etat des \'electrons du syst\`eme est d\'ecrit par une matrice densit\'e, c'est \`a dire un op\'erateur auto-adjoint \`a trace sur $L ^2 (\R ^3)$. Ceci est une description plus compl\`ete que celle bas\'ee uniquement sur une densit\'e \'electronique comme dans la section pr\'ec\'edente, mais il y a cependant une perte d'information, correspondant \`a deux hypoth\`eses de mod\'elisations indiqu\'ees plus bas.

On imagine un syst\`eme \`a $N$ \'electrons d\'ecrit par une fonction d'onde $\Psi \in L ^2 (\R ^{3N})$ satisfaisant le principe de Pauli \eqref{eq:antisym}. Puisqu'on veut d\'ecrire un syst\`eme infini il faudrait pouvoir prendre $N=\infty$, ce qui n'a pas de sens math\'ematiquement \`a ce niveau puisque l'espace lui-m\^eme d\'epend de $N$. Il se trouve cependant que l'\'energie d'un syst\`eme quantique \`a $N$-corps avec interactions de paire ne d\'epend que de deux op\'erateurs li\'es \`a la fonction d'onde $\Psi$: les matrices densit\'e \`a un et deux corps d\'efinies par leurs noyaux
\begin{align}\label{eq:matrices}
\gamma_1 (x;y) :=& N \int_{\R ^{3(N-1)}} \overline{\Psi(x,x_2,\ldots,x_N)} \Psi(y,x_2,\ldots,x_N)dx_2\ldots dx_N\nonumber\\
\gamma_2 (x,x';y,y') :=& N(N-1) \int_{\R ^{3(N-2)}} \overline{\Psi(x,x',x_3,\ldots,x_N)} \Psi(y,y',x_3,\ldots,x_N) dx_3\ldots dx_N.
\end{align}
Les matrices densit\'es $\gamma_1$ et $\gamma_2$ sont des op\'erateurs \`a trace agissant respectivement sur $L ^2 (\R ^3)$ et $L ^2 (\R ^3) ^{\otimes 2} \simeq L ^2 (\R ^6)$. Ces espaces ne d\'ependent plus de $N$, la d\'ependance en le nombre de particules $N$ \'etant simplement encod\'ee dans les traces 
\[
\tr (\gamma_1) = N,\quad \tr(\gamma_2) = N(N-1), 
\]
que l'on peut facilement normaliser. Ces objets, contrairement \`a la fonction d'onde,  sont donc bien adapt\'es pour \'etudier la limite $N\to \infty$. On a, par exemple dans le cas du Hamiltonien \eqref{eq:schro N corps}, la formule 
\begin{equation}\label{eq:ener schro matr}
\bral \Psi, H_N \Psi \ketr = \Tr_{L ^2 (\R ^3)} \left(-\frac{1}{2} \Delta \gamma_1 \right) + \frac{1}{2} \Tr_{L ^2 (\R ^6)} \left( \frac{1}{|x-y|} \gamma_2 \right)
\end{equation}
pour l'\'energie d'une fonction d'onde $\Psi \in L ^2 (\R ^{3N})$ en fonction des matrices densit\'e, qui permet de formuler le probl\`eme uniquement en termes de $\gamma_1$ et $\gamma_2$.

\medskip

Notre description des \'electrons du cristal (d\'esign\'es sous le vocable \'evocateur de ``mer de Fermi'') est bas\'ee sur deux hypoth\`eses suppl\'ementaires:

\medskip

(H5) \emph{Les \'electrons du cristal ont les corr\'elations minimales permises par le principe de Pauli.
}

(H6) \emph{L'\'energie dite d'\'echange des \'electrons du cristal est n\'egligeable. 
}
\medskip

Des particules compl\`etement d\'ecorr\'ell\'ees auraient une fonction d'onde
$$\Psi (x_1,\ldots,x_N)= \psi(x_1) \ldots \psi(x_N)$$
avec $\psi \in L ^2 (\R ^3) $, ce qui revient \`a dire que toutes les particules sont dans l'\'etat quantique $\psi$. Une telle fonction d'onde n'est pas antisym\'etrique et donc incompatible avec le principe de Pauli \eqref{eq:antisym} qui interdit \`a deux \'electrons d'occuper simultan\'ement le m\^eme \'etat. La fonction d'onde la plus simple et la moins corr\'elée que l'on puisse imaginer satisfaisant \eqref{eq:antisym} est un d\'eterminant de Slater
\begin{equation}\label{eq:Slater}
\Psi (x_1,\ldots,x_N)  = \det \left( \psi_i (x_j) \right)
\end{equation}
o\`u les orbitales $\psi_1,\ldots,\psi_N\in L ^2 (\R ^3)$ d\'ecrivent l'\'etat des $N$ \'electrons. L'hypoth\`ese~5 correspond \`a supposer que les \'electrons peuvent \^etre d\'ecrits par un d\'eterminant de Slater. Dans le langage des matrices densit\'e, plus adapt\'e \`a la limite thermodynamique $N\to \infty$, ceci revient \`a demander que la matrice densit\'e \`a deux corps soit enti\`erement d\'etermin\'ee par la matrice \`a un corps:
\begin{equation}\label{eq:2 corps Slater}
\gamma_2 = \gamma_1 ^{\otimes 2}\left( \one_2 - \Ex \right) 
\end{equation}
avec $\one_2$ l'identit\'e sur $L ^2 (\R ^3)^{\otimes 2} $ et $\Ex$ l'op\'erateur d'\'echange d\'efini par son action sur les produits tensoriels purs
\begin{equation}\label{eq:exchange}
\Ex (u\otimes v)= v\otimes u,\quad \forall u,v\in L ^2 (\R ^3). 
\end{equation}
L'\'energie d'un d\'eterminant de Slater \eqref{eq:Slater} s'exprime donc par la formule \eqref{eq:ener schro matr} uniquement en fonction de sa matrice densit\'e \`a un corps, un op\'erateur \`a trace positif et auto-adjoint sur $L ^2 (\R ^3)$ qui peut s'\'ecrire 
\begin{equation}\label{eq:matrice Slater}
\gamma = \sum_{i=1} ^N \ketl \psi_i \ketr \bral \psi_i \brar 
\end{equation}
o\`u $\ketl \psi_i \ketr \bral \psi_i \brar$ d\'esigne le projecteur orthogonal sur l'espace vectoriel engendr\'e par $\psi_i$. Dans ce langage, le principe de Pauli \eqref{eq:antisym} se traduit par la contrainte 
\begin{equation}\label{eq:Pauli dens mat}
0 \leq \gamma \leq \one
\end{equation}
au sens des op\'erateurs, o\`u on a not\'e $\one$ l'op\'erateur identit\'e sur $L ^2 (\R ^3)$. Notons que nous minimiserons une fonctionnelle de $\gamma$ sur tout l'espace des op\'erateurs auto-adjoints \`a trace satisfaisant \eqref{eq:Pauli dens mat}. Il n'est pas \'evident a priori que l'\'energie minimum dans cette classe soit donn\'ee par un d\'eterminant de Slater, i.e. par un $\gamma$ de forme \eqref{eq:matrice Slater}. C'est cependant vrai \cite{Lieb-81}, ce qui garantit qu'il n'y a pas de perte d'information lorsque nous consid\'erons des matrices densit\'e g\'en\'erales.

\medskip

L'hypoth\`ese 6 postule que le terme d'\'energie d\^u \`a l'op\'erateur d'\'echange dans \eqref{eq:2 corps Slater} est n\'egligeable, ce qui revient \`a supposer que la matrice \`a deux corps (en fonction de laquelle s'exprime l'\'energie d'interaction) est approximativement donn\'ee en fonction de $\gamma$ par 
\[
\gamma_2 = \gamma ^{\otimes 2}, 
\]
l'\'energie de $\gamma$ \'etant obtenue en ins\'erant ce postulat dans la formule \eqref{eq:ener schro matr}. Dans ce cadre, le fait que nous consid\'erons un syst\`eme de fermions n'appara\^it plus que dans la contrainte \eqref{eq:Pauli dens mat}, et nous avons l'immense avantage que l'\'energie des \'electrons est une fonction convexe de la matrice densit\'e. Dans notre vocabulaire (introduit dans \cite{Solovej-91}), la th\'eorie obtenue apr\`es cette simplification sera dite de Hartree-Fock r\'eduite. On l'appelle souvent th\'eorie de Hartree dans la litt\'erature physique, ce qui est l\'eg\`erement impropre puisque la th\'eorie de Hartree correspond stricto sensu \`a consid\'erer la m\^eme fonctionnelle d'\'energie mais sans la contrainte cruciale \eqref{eq:Pauli dens mat}, auquel cas c'est un mod\`ele valable pour des bosons mais pas pour des fermions. 

Les hypoth\`eses 5 et 6 sont parfaitement raisonnables d'un point de vue physique: il est connu que pour un grand syst\`eme fermionique, la limite $N\to \infty$ peut-\^etre calcul\'ee au premier ordre dans une th\'eorie Hartree-Fock r\'eduite (voir les travaux  \cite{LieSim-77b,Lieb-81b,BarGolGotMau-03,BenPorSch-13} et leurs r\'ef\'erences). En fait, on peut m\^eme dans certains cas la calculer en utilisant des th\'eories encore plus simples (de type Thomas-Fermi), et on pourrait donc imaginer utiliser un mod\`ele de ce type pour d\'ecrire les \'electrons du cristal. Remarquons cependant que ce mod\`ele aurait des d\'efauts importants \cite{CanEhr-11}, ce qui constitue une motivation pour ne pas aller trop loin dans la simplification et en rester \`a un mod\`ele de type Hartree-Fock. Nous n\'egligerons le terme d'\'echange pour obtenir une th\'eorie de Hartree-Fock r\'eduite, ceci car \`a cause de son manque de convexit\'e il n'est pas clair du tout (et en tout \'etat de cause, non prouv\'e \`a ce jour) que la fonctionnelle de Hartree-Fock compl\`ete se comporte convenablement dans la limite $N\to \infty$.

\subsection{Mod\`ele rHF pour le cristal parfait}

Nous entrons maintenant plus dans les d\'etails en d\'ecrivant le cadre math\'ematique appropri\'e au cas d'un cristal parfait, avec un r\'eseau cristallin r\'egulier. 

Etant donn\'ee une densit\'e de charge p\'eriodique $\munucper$ pour les noyaux, il s'agit de donner un sens \`a l'\'energie
\begin{equation}\label{eq:cri per formel}
\rhff_{\munucper} [\gamma] = \Tr \left(- \frac{1}{2}\Delta \gamma\right) - \iint_{\R^3 \times \R^3} 
\frac{\rhogam (x)\munucper (y)}{|x-y|}dxdy + \iint_{\R^3 \times \R^3} 
\frac{\rhogam (x)\rhogam (y)}{|x-y|}dxdy,
\end{equation}
o\`u $\gamma$ est une matrice densit\'e \`a un corps, $\rhogam$ est la densit\'e de charge correspondante, donn\'ee formellement par $\rhogam(x) = \gamma (x,x)$ et plus rigoureusement par 
\begin{equation}\label{eq:rhogam}
\rhogam (x) = \sum_{i=1} ^N |\psi_i (x)| ^2  
\end{equation}
si $\gamma$ est de la forme \eqref{eq:matrice Slater}. Par rapport \`a \eqref{eq:ener schro matr} nous avons inclut l'interaction \'electrostatique entre les noyaux et les \'electrons (deuxi\`eme terme) et nous avons exprim\'e l'interaction \'electrons-\'electrons en utilisant la densit\'e de charge (diagonale de la matrice densit\'e). Notons l'absence d'un terme d'\'echange qui serait de la forme
\[
- \iint_{\R^3 \times \R^3} \frac{\left|\gamma(x,y)\right| ^2}{|x-y|}dxdy, 
\]
et rendrait la fonctionnelle non convexe.

Pour un syst\`eme $\cL$-p\'eriodique, l'espace variationnel appropri\'e est celui des matrices commutant avec les translations du r\'eseau (qui correspondent \`a des fonctions d'onde invariantes par les translations du r\'eseau):
\begin{equation}\label{eq:etats parfaits}
\cP_{\rm per} := \left\{ \gamma \in \cB \left( L ^2 (\R ^3) \right)| \; \gamma = \gamma ^*,\; 0\leq \gamma \leq 1,\; \gamma \tau_{k} =  \tau_{k} \gamma  \; \forall k\in \cL \right\}. 
\end{equation}
Ici et dans la suite nous notons $\cB (\gH)$ l'espace des op\'erateurs born\'es sur un espace de Hilbert $\gH$. Comme not\'e pr\'ec\'edemment, les matrices densit\'e admissibles sont autoadjointes et satisfont \eqref{eq:Pauli dens mat}.

La bonne fa\c{c}on de donner un sens rigoureux \`a l'\'energie par unit\'e de volume correspondant \`a \eqref{eq:cri per formel} est de prendre avantage de l'invariance par translation du probl\`eme en utilisant une d\'ecomposition en ondes de Bloch \cite{ReeSim4} des op\'erateurs $\gamma\in \cP_{\rm per}$ admissibles.  En notant $\Gamma ^* = 2\pi \Gamma$ la zone de Brillouin du r\'eseau, tout op\'erateur de $\cP_{\rm per}$ admet une d\'ecomposition sous la forme 
\begin{align}\label{eq:Bloch decomp}
 \gamma &= \frac{1}{(2\pi) ^3} \int_{\Gamma ^*} \gamma_{\xi} d\xi, \; \gamma_{\xi} \in \cB(L ^2 _{\xi} (\Gamma))\nonumber\\
 L ^2 _{\xi} (\Gamma)&= \left\{ u\in L ^2 _{\rm loc} (\R ^3),\; \tau_k u = e ^{-ik \cdot \xi} u \; \forall k\in \cL \right\}
\end{align}
au sens o\`u toute fonction de $L ^2 (\R ^3)$ s'\'ecrit
\[
u =  \frac{1}{(2\pi) ^3} \int_{\Gamma ^*} u_{\xi} d\xi , \; u_{\xi} \in L ^2 _{\xi} (\Gamma)
\]
et $\gamma \in \cP_{\rm per}$ agit comme 
\[
 \gamma u = \frac{1}{(2\pi) ^3} \int_{\Gamma ^*} \gamma_{\xi} u_{\xi} d\xi. 
\]
Cette repr\'esentation correspond \`a la d\'ecomposition en fibr\'es $L ^2 (\R ^3) = \int_{\Gamma ^* } ^{\oplus} L ^2 _{\xi} (\Gamma)$. 

Le sens pr\'ecis \`a donner \`a \eqref{eq:cri per formel} est alors 
\begin{equation}\label{eq:cri per}
\rhff_{\munucper} [\gamma] = \frac{1}{(2\pi) ^3} \int_{\Gamma ^*} \Tr \left(- \frac{1}{2}\Delta_{\xi} \gamma_{\xi} \right) d\xi + D_{G}\left(\munucper - \rhogam,\munucper - \rhogam\right)
\end{equation}
o\`u $\rhogam$ se d\'efinit (formellement, voir-ci dessus) comme 
\[
\rhogam(x) = \frac{1}{(2\pi) ^3} \int_{\Gamma ^*} \gamma_{\xi} (x,x) d\xi 
\]
et $D_G (.,.)$ est l'interaction Coulombienne entre densit\'es de charge p\'eriodiques:
\begin{equation}\label{eq:Coul per}
D_{G}(f,g) = \iint_{\Gamma \times \Gamma} f(x) G(x,y) g(y) dxdy 
\end{equation}
avec $G$ la fonction de Green d\'efinie par 
\[
 \begin{cases}
  -\Delta G = 4\pi \left( \sum_{k\in \cL} \delta_k - 1 \right) \mbox{ in } \R ^3 \\
  G \;\cL-\mbox{p\'eriodique},\; \min_{\R ^3} G = 0.
 \end{cases}
\]
Remarquons que l'\'energie \eqref{eq:cri per} inclut le terme constant $D(\munucper,\munucper)$. 

L'\'etat fondamental du cristal parfait p\'eriodique est d\'efini par le probl\`eme de minimisation 
\begin{equation}\label{eq:defi crist parfait}
\rhfe_{\munucper} = \inf \left\{ \rhff_{\munucper} [\gamma],\; \gamma  \in \cP_{\rm per},\; \int_{\Gamma} \rhogam = Z \right\} 
\end{equation}
o\`u nous notons 
\[
 Z = \int_{\Gamma} \munucper
\]
et la minimisation s'effectue donc sous une contrainte de neutralit\'e locale de la m\^eme forme que \eqref{eq:neutral local}. Le choix pour l'espace variationnel \`a consid\'erer pour d\'efinir les termes de \eqref{eq:cri per} est \'evident, il faut notamment que $\gamma_{\xi}$ et $|\nabla| \gamma_{\xi} |\nabla|$ soient des op\'erateurs \`a trace pour tout $\xi$, ce qui garantit l'existence d'une densit\'e de charge associ\'ee et permet de donner un sens \`a tous les termes de la fonctionnelle.

L'existence d'un minimiseur est \'etablie dans \cite{CatBriLio-01}, son unicit\'e dans \cite{CanDelLew-08a}. La justification du mod\`ele par un argument de limite thermodynamique est \'egalement pr\'esent\'ee dans \cite{CatBriLio-01}. Il sera crucial pour la suite de savoir que l'unique minimiseur $\FSm$ satisfait l'\'equation d'Euler-Lagrange (voir \cite{CanDelLew-08a})
\begin{equation}\label{eq:EEL crist parf}
\begin{cases}
\displaystyle \FSm = \oneep (\FSh)\\
\displaystyle \FSh = -\frac{1}{2} \Delta + \FSp \\
\displaystyle \FSp = \left(\munucper - \FSd \right) \star \coulker 
\end{cases}
\end{equation}
avec $\FSd (x) = \FSm (x,x)$ la densit\'e associ\'ee et $\FSl$ le multiplicateur de Lagrange, appel\'e niveau de Fermi dans ce contexte. Ici $\oneep (\FSh)$ d\'esigne le projecteur spectral jusqu'au niveau $\FSl$ associ\'e \`a $\FSh$, que l'on d\'efinit ais\'ement \cite{ReeSim4} en diagonalisant chaque composante de Bloch $(\FSh)_{\xi}$ de $\FSh$ qui est un op\'erateur de Schr\"odinger p\'eriodique. L'interpr\'etation de \eqref{eq:EEL crist parf} est fort naturelle: \`a cause du principe de Pauli, des \'electrons libres (sans interaction), remplissent les niveaux d'\'energie successifs de leur Hamiltonien. Ici les interactions sont prises en compte donc les \'electrons du cristal parfait remplissent les niveaux d'\'energie d'un Hamiltonien auto-consistant (dit de champ moyen) $\FSh$ qui est le Hamiltonien effectif ressenti par chaque \'electron. Le niveau de remplissage est d\'etermin\'e par le niveau de Fermi $\FSl$ pour assurer la neutralit\'e locale du cristal.

\subsection{Mod\`ele rHF pour le cristal avec d\'efaut}

Nous d\'ecrivons maintenant le mod\`ele pour cristal avec d\'efaut introduit dans \cite{CanDelLew-08a,CanDelLew-08b}. L'id\'ee principale est de d\'efinir l'\'energie du cristal en pr\'esence d'un d\'efaut de charge local $\nu \in L ^1 (\R ^3)$ par r\'ef\'erence au cristal parfait. On obtient ainsi la diff\'erence d'\'energie avec la mer de Fermi p\'eriodique caus\'e par la pr\'esence du d\'efaut $\nu$, une m\'ethodologie inspir\'ee des travaux \cite{ChaIra-89,ChaIraLio-89,HaiLewSer-05a,HaiLewSer-05b,HaiLewSer-09} o\`u l'\'energie de particules relativistes en pr\'esence d'une mer de Dirac d'\'electrons virtuels est rigoureusement d\'efinie, voir \cite{HaiLewSerSol-07,Lewin_proc-12} pour une description synth\'etique. Dans le cadre des cristaux le d\'efaut de charge  peut mod\'eliser par exemple une impuret\'e, un noyau manquant ou en exc\`es dans le r\'eseau ou un r\'earrangement local du r\'eseau. Pour la description des polarons, $\nu$ repr\'esentera la densit\'e de charge des \'electrons ajout\'es au cristal.  

\medskip

On \'ecrira l'\'etat $\gamma$ du cristal en pr\'esence du d\'efaut comme une perturbation du cristal parfait:
\[
\gamma = \FSm + Q 
\]
et on exprimera l'\'energie en fonction de $Q$ en faisant le calcul formel
\begin{equation}\label{eq:cri def formel}
 \rhff_{\munucper+\nu}[\gamma] - \rhff_{\munucper+\nu}[\FSm] = \Tr \left(  \FSh   Q \right) + D(\nu,\rhoQ) +\frac{1}{2} D (\rhoQ,\rhoQ)
\end{equation}
avec 
\begin{equation}\label{eq:Coul}
D(\nu,\mu) = \int_{\R ^3} \nu \left(\mu \star \coulker\right) 
\end{equation}
l'interaction Coulombienne et $\rhoQ$ la densit\'e de charge associ\'ee \`a $Q$, dans l'esprit de \eqref{eq:rhogam}. Bien que les deux termes au membre du gauche soient infinis, on peut donner un sens rigoureux au membre de droite, et d\'emontrer que c'est bien cette quantit\'e qui appara\^it \`a la limite thermodynamique si on part d'un mod\`ele avec un nombre fini de particules pour lequel le membre de gauche fait sens et peut se calculer. Il s'agit en fait de calculer le deuxi\`eme ordre dans la limite thermodynamique en pr\'esence d'un d\'efaut localis\'e, le premier \'etant donn\'e (par unit\'e de volume) par la fonctionnelle p\'eriodique d\'ecrite \`a la section pr\'ec\'edente.

Ici nous pr\'esentons le cadre fonctionnel adapt\'e \`a la d\'efinition et l'\'etude de \eqref{eq:cri def formel}, renvoyant \`a \cite{CanDelLew-08a,CanLew-10} pour les d\'etails et l'\'etude de la limite thermodynamique correspondante. Le point le plus important est sans aucun doute la n\'ecessit\'e de revoir le sens du symbole $\Tr$ dans \eqref{eq:cri per formel}. Il convient en fait de g\'en\'eraliser la notion de trace, car la perturbation $Q$ n'est pas a priori un op\'erateur \`a trace (a posteriori non plus d'ailleurs, voir plus bas). Suivant une id\'ee d\'evelopp\'ee dans \cite{BacBarHelSie-99,HaiLewSer-05a,HaiLewSer-05b}, on d\'ecompose $Q$ sous la forme 
\begin{align}\label{eq:decomp trace}
 Q &= \FSm Q \FSm + (1-\FSm) Q (1-\FSm) + \FSm Q (1-\FSm) + (1-\FSm) Q \FSm \nonumber \\
 &= \Qmm + \Qpp + \Qmp + \Qpm
\end{align}
et on consid\'erera des op\'erateurs pour lesquels les composantes diagonales $\Qmm$ et $\Qpp$ sont s\'epar\'ement \`a trace. On peut alors d\'efinir la trace g\'en\'eralis\'ee
\begin{equation}\label{eq:gene trace}
 \Tro (Q) := \Tr (\Qmm) + \Tr(\Qpp). 
\end{equation}
Puisque, d'apr\`es \eqref{eq:EEL crist parf}, $\FSm$ est un projecteur orthogonal, il est \'evident que $\Tr$ et $\Tro$ co\"incident pour des op\'erateurs \`a trace. Cependant un op\'erateur peut avoir une trace g\'en\'eralis\'ee sans avoir de trace au sens usuel: on a fix\'e une d\'ecomposition de l'espace de Hilbert ambiant et calcul\'e la trace dans une base associ\'ee \`a cette d\'ecomposition, ce qui ne garantit en rien que la trace puisse se calculer avec une autre d\'ecomposition et donner le m\^eme r\'esultat. 

Il est aussi important de noter que le principe de Pauli implique la contrainte $0 \leq \gamma \leq 1$, i.e.
\begin{equation}\label{eq:Pauli def}
-\FSm \leq Q \leq 1-\FSm 
\end{equation}
au sens des op\'erateurs, et que ceci est en fait \'equivalent \`a la contrainte
\begin{equation}\label{eq:Pauli def 2}
 \Qpp - \Qmm \geq Q ^2.
\end{equation}
Avec ces \'el\'ements en main, le sens naturel \`a donner au terme d'\'energie cin\'etique de \eqref{eq:cri def formel} est 
\[
\Tro \left(  \FSh   Q \right) := \Tr \FSh \Qpp + \Tr \FSh \Qmm = \Tr \left| \FSh-\FSl \right| \left( \Qpp - \Qmm \right) + \FSl \Tro Q.  
\]
On a alors la propri\'et\'e de coercivit\'e 
\begin{equation}\label{eq:coer}
\Tro \left(  \FSh -\FSl  \right) Q  \geq  \Tr \left| \FSh-\FSl \right| Q ^2 
\end{equation}
qui permet de d\'efinir un espace variationnel naturel et de d\'eduire de bonnes propri\'et\'es, telle que l'existence d'une densit\'e de charge pour les $Q$ admissibles. Sans plus de d\'etail (voir \cite{CanDelLew-08a}), donnons l'expression  
\begin{equation}\label{eq:crys def func}
\crysf [\nu,Q] = \Tro \left( \left( \FSh - \FSl \right) Q \right) + D(\nu,\rhoQ) +\frac{1}{2} D (\rhoQ,\rhoQ)
\end{equation}
pour l'\'energie du cristal en pr\'esence du d\'efaut $\nu$. On a utilis\'e le niveau de Fermi $\FSl$ du cristal parfait comme potentiel chimique pour fixer une r\'ef\'erence d'\'energie et donner un sens au terme d'\'energie cin\'etique. Pour que la notion de cristal perturb\'e fasse sens il faut faire l'hypoth\`ese  

\medskip

\noindent (H7) \emph{Le cristal h\^ote est un isolant. L'op\'erateur de Schr\"odinger p\'eriodique $\FSh$ d\'efini par \eqref{eq:EEL crist parf} a un trou spectral entre sa $Z$-i\`eme et $(Z+1)$-i\`eme bande, et $\FSl$ est dans ce gap.}

\medskip

On peut alors d\'efinir le changement d'\'energie minimal en pr\'esence du d\'efaut $\nu$ comme   
\begin{equation}\label{eq:crys def ener}
\inf_{- \FSm \leq Q \leq 1 -\FSm} \left\{ \Tro \left( \left( \FSh - \FSl \right) Q \right) + D(\nu,\rhoQ) +\frac{1}{2} D (\rhoQ,\rhoQ) \right\}.
\end{equation}
Au vu de la propri\'et\'e de coercivit\'e \eqref{eq:coer}, l'espace variationnel adapt\'e \`a la minimisation est\footnote{Le lecteur doit avoir en t\^ete la correspondance $|\FSh - \FSl| \simeq -\Delta = |\nabla| ^2$.} 
\begin{equation}\label{eq:vari space crys}
 \Q = \left\lbrace |\nabla | Q \in \Sch ^2,\:Q = Q^* ,\: |\nabla| \Qpp |\nabla| \in \Sch ^1, \: |\nabla| \Qmm |\nabla| \in \Sch ^1\right\rbrace
\end{equation}
o\`u les classes de Schatten sont donn\'ees par 
\[
\Sch ^p (L^2 (\R^3)):= \left\lbrace Q \in \cB\left( L^2 (\R^3)\right),\: \Tr \left( |Q| ^p \right) ^{1/p} < +\infty \right\rbrace. 
\]
En particulier, $\gS ^1$ est l'espace des op\'erateurs \`a trace et $\gS ^2$ l'espace des op\'erateurs Hilbert-Schmidt. Bien s\^ur il faut aussi donner un sens \`a la densit\'e $\rhoQ$ dans \eqref{eq:crys def func}, ce qui se fait par dualit\'e: pour tout $Q\in \Q$, il existe $\rhoQ \in L^2 (\R ^3)$) tel que
\[
 \Tro (Q V) = \int_{\R ^3} \rhoQ V
\]
pour tout $V$ assez r\'egulier, et on a $D(\rhoQ,\rhoQ) < \infty$.

L'existence d'un minimiseur pour \eqref{eq:crys def func} dans la classe \eqref{eq:vari space crys} est \'etablie dans \cite{CanDelLew-08a} et certaines de ses propri\'et\'es importantes sont fournies dans \cite{CanLew-10}. En particulier il est \'etabli que le cadre fonctionnel plut\^ot complexe \'evoqu\'e ci-dessus ne peut pas se simplifier: un minimiseur n'est en g\'en\'eral (c'est-\`a-dire \`a part \'eventuellement pour le r\'eseau cubique $\cL= \Z ^3$) pas un op\'erateur \`a trace. Ceci est une propri\'et\'e physique fondamentale due \`a la nature des interactions Coulombiennes: la r\'eponse du cristal \`a un d\'efaut de charge comporte des oscillations \`a longue port\'ee, ce qui se traduit par le fait que la densit\'e de charge $\rhoQ$ de la perturbation du cristal n'est pas dans $L ^1 (\R ^3)$. Il est donc a posteriori logique que la d\'efinition du mod\`ele soit assez d\'elicate.

\section{Un nouveau mod\`ele pour le (petit) polaron}\label{sec:model}

Nous pouvons maintenant d\'efinir proprement le mod\`ele que nous avons \'etudi\'e \cite{LewRou-11,LewRou-12}. Dans l'esprit de la pr\'esentation g\'en\'erale donn\'ee auparavant l'\'energie d'un seul polaron est donn\'ee par
\begin{equation}
\cE[\psi]=\frac1{2}\int_{\R^3}|\nabla\psi(x)|^2\,dx+\int_{\R^3}V^0_{\rm per}(x)|\psi(x)|^2\,dx+F_{\rm crys}\big[|\psi|^2\big]
\label{eq:model-intro1}
\end{equation}
avec $\psi$ la fonction d'onde du polaron, et l'\'energie de $N$ polarons en interaction par 
\begin{multline}\label{eq:model-introN}
\cE[\Psi] = \int_{\R^{3N}}\left(\frac12\sum_{j=1}^N \left|\nabla_{x_j}\Psi(x_1,...,x_N)\right|^2+\sum_{1\leq k<\ell\leq N}\frac{|\Psi(x_1,...,x_N)|^2}{|x_k-x_\ell|}\right)dx_1\cdots dx_N  \\
+\int_{\R ^3} \FSp \rhoP + \cryse [\rhoP]
\end{multline}
o\`u $\Psi \in L ^2 (\R ^{3N})$ est la fonction d'onde du syst\`eme de $N$ polarons, auquel on associe une densit\'e de charge $\rhoP$ par la formule \eqref{eq:rhoP}. 

Le potentiel $\FSp$ est celui g\'en\'er\'e par le cristal au repos \eqref{eq:EEL crist parf} et la fonctionnelle $\cryse$ est d\'efinie \`a partir du mod\`ele pour le cristal avec d\'efaut de charge de la section pr\'ec\'edente:
\begin{equation}\label{eq:crys def ener}
\cryse[\nu] = \inf_{- \FSm \leq Q \leq 1 -\FSm} \left\{ \Tro \left( \left( \FSh - \FSl \right) Q \right) + D(\nu,\rhoQ) +\frac{1}{2} D (\rhoQ,\rhoQ) \right\}.
\end{equation}
Faisons quelques commentaires avant de commencer l'\'etude math\'ematique:
\begin{itemize}
\item Ce mod\`ele peut \^etre mis sous la forme g\'en\'erale d\'ecrite \`a la Section \ref{sec:generic} en utilisant une formulation de type fonctionnelle de densit\'e~\cite{Lieb-83b}:
avec $\delta\rho=\rho_Q$, on d\'efinit l'\'energie pour cr\'eer la perturbation $\delta\rho$ dans le cristal par
$$\cF^{\,'}_{\rm crys}[\delta\rho]=\inf_{\substack{-\gamma^0_{\rm per}\leq Q\leq 1-\gamma^0_{\rm per}\\ \rho_Q=\delta\rho}} \left(\Tro \left( \left( \FSh - \FSl \right) Q \right) +\frac{1}{2} D (\rhoQ,\rhoQ) \right),$$
et $F_{\rm crys}[\nu]$ peut aussi s'exprimer
$$F_{\rm crys}[\nu]=\inf_{\delta\rho\geq-\rho^0_{\rm per}}\Big(D(\nu,\delta\rho)+\cF^{\,'}_{\rm crys}[\delta\rho]\Big).$$
Cette formulation n'est pas vraiment utile pour la suite, et ne devrait pas faire oublier le fait que notre mod\`ele de cristal inclut l'effet du principe de Pauli (sous la forme de la contrainte impos\'ee \`a $Q$ dans la minimisation), contrairement aux fonctionnelles plus simples \'evoqu\'ees auparavant.
\item La nouveaut\'e principale de ce mod\`ele est qu'il est appropri\'e pour la description d'un petit polaron, c'est-\`a-dire que nous n'avons fait aucune hypoth\`ese sur la taille relative du polaron et de la cellule unit\'e du r\'eseau cristallin. Les mod\`eles les plus utilis\'es dans la litt\'erature, i.e. ceux de Pekar et de Fr\"ohlich sont tous deux limit\'es au grand polaron: ils ne prennent pas en compte la structure d\'etall\'ee du cristal et ils postulent donc que la fonction d'onde $\psi$ est \'etal\'ee sur une \'echelle de longueur tr\`es grande devant la longueur caract\'eristique du cristal.
\item Le prix \`a payer pour cet avantage est le caract\`ere hautement non lin\'eaire de la fonctionnelle que nous avons introduite, qui pose des difficult\'es math\'ematiques consid\'erables pour obtenir l'existence d'\'etats li\'es, comme expliqu\'e plus bas.  
\item Ce mod\`ele n\'eglige les corr\'elations polarons-cristal, il est donc de nature assez diff\'erente du mod\`ele de Fr\"ohlich \'evoqu\'e auparavant. C'est en revanche une g\'en\'eralisation naturelle du mod\`ele de Pekar, et nous verrons \`a la Section \ref{sec:deriv Pekar} qu'il permet une d\'erivation rigoureuse des fonctionnelles \eqref{eq:Pekar 1} et \eqref{eq:Pekar N}.
\end{itemize}

La suite de cette section est consacr\'ee \`a l'existence d'\'etats li\'es. Nous r\'esumons le travail \cite{LewRou-12} o\`u l'\'equivalent des th\'eor\`emes \ref{theo:PTg1} et \ref{theo:PTgN} pour les fonctionnelles \eqref{eq:model-intro1} et \eqref{eq:model-introN} est d\'emontr\'e.

\subsection{Existence d'\'etats li\'es}

L'\'etat fondamental d'un syst\`eme de petits polarons est d\'efini par la minimisation des fonctionnelles \eqref{eq:model-intro1} et \eqref{eq:model-introN}: 
\begin{align}
E(1): =& \inf \left\lbrace \E [\psi],\: \psi \in H ^1 (\R ^{3}),\ \int_{\R ^{3} } |\psi| ^2 = 1 \right\rbrace 
\label{eq:energyN}\\
E(N): =& \inf \left\lbrace \E [\Psi],\: \Psi \in H ^1 (\R ^{3N}), \: \Psi \mbox{ satisfait \eqref{eq:antisym}, } \int_{\R ^{3N} } |\Psi| ^2 = 1 \right\rbrace. 
\end{align}

Dans le cas d'un seul polaron nous avons le r\'esultat suivant:

\begin{theorem}[\textbf{Existence de petits polarons} \cite{LewRou-12}]\label{theo:E_1}\ \\
Pour $N=1$, on a  
\begin{equation}
E(1)<E_{\rm per}:=\inf\sigma\left(-\frac1{2m}\Delta+V^0_{\rm per}\right).
\label{eq:E_1} 
\end{equation}
Toutes les suites minimisantes pour $E(1)$ convergent fortement dans $H^1(\R^3)$ vers un minimiseur, \`a extraction et translation pr\`es.
\end{theorem}

En d'autres termes, l'existence d'\'etats li\'es pour un polaron est garantie par le mod\`ele \eqref{eq:model-intro1}. L'in\'egalit\'e \eqref{eq:E_1} exprime le fait que l'\'energie d'un petit polaron est toujours plus petite que l'\'energie d'un \'electron libre dans le potentiel $\FSp$, i.e. qu'il est plus favorable d'avoir un \'etat li\'e qu'un \'etat diffusif. Notons que que la n\'ecessit\'e d'une translation pour rendre les suites minimisantes pr\'e-compactes est naturelle vu que le mod\`ele est invariant par toutes les translations du r\'eseau cristallin.

Le cas de $N\geq 2$ polarons est plus subtil: vu le Th\'eor\`eme \ref{thm:Pekar N non exist} pour la fonctionnelle de Pekar, on ne peut s'attendre \`a avoir un \'etat li\'e quelle que soit la valeur de $N$. L'existence d'un minimiseur est par contre garantie si certaines in\'egalit\'es de liaison sont v\'erifi\'ees: 

\begin{theorem}[\textbf{Th\'eor\`eme HVZ pour le petit $N$-polaron} \cite{LewRou-12}]\label{theo:HVZ}\ \\
Pour $N\geq2$, les propositions suivantes sont \'equivalentes:
\begin{enumerate}
\item On a 
\begin{equation}\label{eq:binding petit}
 E(N) < E(N-k) + E(k) \mbox{ pour tout } k=1,\ldots, N-1.
\end{equation}
\item Toutes les suites minimisantes pour $E(N)$ convergent fortement dans $H^1(\R^{3N})$ vers un minimiseur, \`a extraction et translation pr\`es.
\end{enumerate}
\end{theorem}

Ceci est la g\'en\'eralisation naturelle du th\'eor\`eme HVZ~\cite{Hun-66,VanWinter-64,Zhislin-71} dans le cas des polarons. Les in\'egalit\'es \eqref{eq:binding petit} expriment le fait qu'il n'est pas favorable de s\'eparer le $N$-polaron en un $k$-polaron et un $(N-k)$-polaron, ce pour toute valeur de $0< k < N$. Dans le langage de la concentration-compacit\'e \cite{Lions-84,Lions-84b}, il s'agit d'\'eviter le d\'efaut de compacit\'e d\^u \`a la dichotomie des suites minimisantes. L'évanescence est emp\^ech\'ee par l'in\'egalit\'e \eqref{eq:E_1}. Bien s\^ur, l'in\'egalit\'e large correspondant \`a \eqref{eq:binding petit} est toujours vraie: puisqu'on regarde un syst\`eme dans tout l'espace, on peut toujours construire un \'etat test avec $k$ polarons sur la Lune et $N-k$ polarons sur Terre, et ces deux syst\`eme n'interagiront pratiquement pas, ce qui implique que l'\'energie de l'\'etat test est donn\'ee par le membre de droite de \eqref{eq:binding petit}. On peut toutefois imaginer qu'il y ait existence de minimiseurs dans certains cas limites sans que \eqref{eq:binding petit} soit valable. Seules certaines suites minimisantes seront pr\'ecompactes dans ce cas, voir par exemple \cite{FraLieSei-12} o\`u un cas de ce genre est \'etudi\'e dans le cadre de la th\'eorie de Pekar.  

\medskip

Le Th\'eor\`eme \ref{theo:E_1} se prouve dans un cadre de concentration compacit\'e classique, \`a la Lions. La preuve du Th\'eor\`eme \ref{theo:HVZ} est plus subtile: il s'agit de traiter un probl\`eme \`a $N$ corps non lin\'eaire. Une th\'eorie g\'en\'erale a \'et\'e d\'eveloppée par Mathieu Lewin \cite{Lewin-11} pour g\'erer ce genre de probl\`eme, le cadre propos\'e couvrant \`a la fois le cas du Th\'eor\`eme \ref{theo:PTgN} et (\`a des adaptations mineures pr\`es) celui du Th\'eor\`eme \ref{theo:HVZ}. Il s'agit de combiner les m\'ethodes dites ``g\'eom\'etriques'' invent\'ees pour le probl\`eme \`a $N$ corps dans les ann\'ees 80 (voir par exemple \cite{Sigal-82,Simon-77}) avec l'approche par concentration compacit\'e. La difficult\'e principale vient du fait que les probl\`emes de minimisation d\'efinissant les termes de \eqref{eq:binding petit} sont pos\'es dans des espaces de Hilbert diff\'erents, correspondant \`a des nombres de particules diff\'erents. La m\'ethode g\'eom\'etrique met tous ces probl\`emes sous le m\^eme toit en utilisant l'espace de Fock tronqu\'e $\C \oplus L ^2 (\R ^3) \oplus L ^2 (\R ^6) \oplus \ldots \oplus L ^{2} (\R ^{3N})$. Une notion de convergence dans cet espace est introduite, adapt\'ee aux th\'eor\`emes de type HVZ. En particulier, une suite de fonctions $\Psi_n \in L ^2 (\R ^{3N})$, vue comme un \'el\'ement de l'espace de Fock tronqu\'e, pourra converger vers une fonction $\Psi \in L ^2 (\R^{3M})$ avec $M<N$. Un exemple important, typique des difficult\'es que l'on rencontre dans ce contexte, est donn\'e par une suite $\Psi_n \in L ^2 (\R ^6)$ avec $$\Psi_n (x,y) = \left(\psi \otimes \phi_n\right) (x,y) = \psi(x) \phi_n (y)$$
o\`u $\phi_n \wto 0$ dans $L ^2 (\R ^3)$. Au sens de la convergence faible usuelle de $L ^2 (\R ^6)$ on a $\Psi \wto 0$, mais au sens de la convergence g\'eom\'etrique d\'efinie dans \cite{Lewin-11} on a $\Psi_n \wto_g \psi$, ce qui permet de capturer des informations plus pr\'ecises sur le comportement de la suite: ici on voit que le d\'efaut de compacit\'e n'est due qu'\`a une des deux variables.  

Un autre outil g\'en\'eralis\'e dans \cite{Lewin-11} est la localisation d'une fonction dans une boule, outil typique de la m\'ethode de concentration compacit\'e. La g\'en\'eralisation naturelle de ce concept dans le cadre des preuves de th\'eor\`emes de type HVZ n\'ecessite d'utiliser l'espace de Fock: la localisation naturelle d'une fonction d'onde $\Psi \in L ^2 (\R ^{3N})$, c'est-\`a-dire l'\'etat quantique construit en localisant les matrices de densit\'e associ\'ees \`a $\Psi$ est en g\'en\'eral un \'etat mixte sur l'espace de Fock tronqu\'e. 

Nous renvoyons \`a \cite{Lewin-11} pour plus de d\'etails sur les m\'ethodes g\'eom\'etriques (le th\'eor\`eme g\'en\'eral que nous utilisons est le Th\'eor\`eme 25 de cet article) et mentionnons dans la suite seulement les probl\`emes sp\'ecifiquement li\'es \`a nos fonctionnelles d'\'energie pour le polaron.

\medskip

La propri\'et\'e cruciale dont on a besoin pour impl\'ementer une approche de type concentration-compacit\'e est celle de ``d\'ecouplage \`a l'infini'': \'etant donn\'ees deux densit\'es de charge bien localis\'ees, et tr\`es \'eloign\'ees l'une de l'autre, l'\'energie de la somme des deux densit\'es doit \^etre quasiment \'egale \`a la somme des \'energies des deux densit\'es prises s\'epar\'ement. C'est cette propri\'et\'e qui permet le lien entre compacit\'e des suites minimisantes et in\'egalit\'es de liaison. Elle est assez \'evidente dans le cas de l'interaction Coulombienne utilis\'ee pour d\'efinir la fonctionnelle de Pekar, mais hautement non triviale pour l'interaction via le cristal \eqref{eq:crys def ener}. Le r\'esultats technique principal de \cite{LewRou-12}, qui permet de conclure la preuve des th\'eor\`emes pr\'ec\'edents est 

\begin{proposition}[\textbf{D\'ecouplage \`a l'infini} \cite{LewRou-12}]\label{pro:decouple}\ \\
Soit $(\rho_n)$ une suite d'\'energie \'electrostatique \eqref{eq:Coul} born\'ee:  
$$D(\rho_n,\rho_n) \leq C$$
ind\'ependamment de $n$. On suppose que $\rho_n\wto\rho$ faiblement au sens o\`u
$$ D (\rho_n,\nu) \to D(\rho,\nu),$$
pour tout $\nu$ tel que $D(\nu,\nu) < \infty$.
Alors
\begin{equation}\label{eq:decouple}
\lim_{n\to \infty} \bigg( \cryse[\rho_n] -  \cryse[\rho] - \cryse[\rho_n-\rho] \bigg) = 0.
\end{equation}
\end{proposition} 

Dans cet \'enonc\'e, il faut imaginer $\rho_n$ comme constitu\'e de deux morceaux de charge, $\rho$ et $\rho_n -\rho$, dont les ``supports'' sont infiniment \'eloign\'es dans la limite $n\to \infty$, ce qui est mat\'erialis\'e par la convergence $\rho_n-\rho \wto 0$. Le r\'esultat dit alors que l'\'energie totale de $\rho_n$ est presque la somme des \'energies des deux morceaux de masse. La difficult\'e pour prouver la Proposition \ref{pro:decouple} est la pr\'esence d'oscillations \`a longue port\'ee dans la r\'eponse de la mer de Fermi \`a un d\'efaut de charge, m\^eme local. A cause de celles-ci il n'est pas \'evident \`a priori que la r\'eponse $Q_n$ \`a un d\'efaut de charge $\rho_n$ comme dans l'\'enonc\'e ci-dessus se d\'ecompose lui-m\^eme en deux morceaux avec des propri\'et\'es similaires, l'un localis\'e autour de $\rho$, l'autre suivant $\rho_n - \rho$. Math\'ematiquement, le probl\`eme est que $\rho_{Q_n}$ n'est pas dans $L ^1 (\R ^3)$, ce qui emp\^eche d'utiliser des estimations na\"ives pour estimer son \'energie Coulomb. Il se trouve que des estimations plus faibles peuvent \^etre impl\'ement\'ees, ce qui permet de conclure la preuve.

Pour montrer que la matrice de densit\'e $Q_n$ associ\'ee \`a $\rho_n$ par la minimisation \eqref{eq:crys def ener} se d\'ecompose bien en deux morceaux, on utilise des op\'erateurs de localisation adapt\'es \`a la d\'ecomposition \eqref{eq:decomp trace}. Étant donn\'ees des fonctions lisses  $\chi_R$ et $\eta_R$ localisant respectivement \`a l'int\'erieur et \`a l'ext\'erieur d'une boule de rayon $R$, on introduit les op\'erateurs (inspir\'es du travail ant\'erieur  \cite{HaiLewSer-09})
\begin{eqnarray}\label{eq:localizations}
X_R &=& \FSm \chi_R \FSm + (1-\FSm) \chi_R (1-\FSm) \nonumber \\
Y_R &=& \FSm \eta_R \FSm + (1-\FSm) \eta_R (1-\FSm)
\end{eqnarray}
qui servent \`a localiser les matrices de densit\'e $Q$ en pr\'eservant la contrainte \eqref{eq:Pauli def}. On a une r\'esolution de l'identit\'e approximative: 
$$X_R ^2 + Y_R ^2 \simeq 1$$
pour $R$ grand, en un pr\'ecis\'e dans le Lemme 3.4 de \cite{LewRou-12}. Le fait que la r\'esolution de l'identit\'e ne soit pas exacte est le prix \`a payer pour avoir la propri\'et\'e bien utile que si $Q_1$ et $Q_2$ satisfont \eqref{eq:Pauli def}, alors $X_R Q_1 X_R + Y_R Q_2 Y_R$ satisfait aussi cette contrainte, ce qui est pr\'ecieux pour construire des \'etats tests admissibles constitu\'es de deux morceaux bien s\'epar\'es.

\section{Le grand polaron: d\'erivation du mod\`ele de Pekar}\label{sec:deriv Pekar}

Dans cette section nous r\'esumons la d\'erivation du mod\`ele de Pekar pr\'esent\'ee en d\'etail dans \cite{LewRou-11}.  

\subsection{Limite macroscopique}

Comme expliqu\'e plus haut, le mod\`ele de Pekar d\'ecrit un ``grand polaron'' \'etal\'e sur une \'echelle grande devant le pas du cristal. Pour d\'eduire cette description \`a partir du mod\`ele introduit \`a la section pr\'ec\'edente, nous introduisons un param\`etre $m\ll 1$ pour quantifier le rapport entre les \'echelles macroscopique et microscopiques: 
\begin{equation}
\cE_m[\psi]:=\frac1{2}\int_{\R^3}|\nabla\psi(x)|^2\,dx+m^{-1}\int_{\R^3}V^0_{\rm per}(x/m)|\psi(x)|^2\,dx+m^{-1}F_{\rm crys}\big[m^{3}|\psi(m\cdot)|^2\big],
\label{totf1}
\end{equation}
avec $V^0_{\rm per}$ et $F_{\rm crys}$ comme pr\'ec\'edemment. L'\'energie fondamentale correspondante est donn\'ee par  
\begin{equation}
E_m(1)=\inf \left\lbrace \cE_m [\psi],\ \int_{\R ^3} |\psi| ^2 = 1  \right\rbrace.
\label{tote1}
\end{equation}
De m\^eme, pour $N\geq2$ l'\'energie du $N$-polaron est d\'efinie par 
\begin{multline}
\cE_{m}[\Psi]=\int_{\R^{3N}}\left(\frac12\sum_{j=1}^N|\nabla_{x_j}\Psi(x_1,...,x_N)|^2+\sum_{1\leq k<\ell\leq N}\frac{|\Psi(x_1,...,x_N)|^2}{|x_k-x_\ell|}\right)dx_1\ldots dx_N\\
+m^{-1}\int_{\R^3}V^0_{\rm per}(x/m)\,\rho_\Psi(x)\,dx+m^{-1}F_{\rm crys}\big[m^{3}|\rhoP(m\cdot)|^2\big],
\label{totfN}
\end{multline}
et l'\'energie fondamentale associ\'ee est 
\begin{equation}\label{toteN}
\tote(N) := \inf \left\lbrace \cE_m [\Psi], \: \int_{\R ^{3N}} |\Psi| ^2 = 1,\: \Psi \mbox{ satisfaisant \eqref{eq:antisym} }  \right\rbrace.
\end{equation}
Dans ces fonctionnelles nous imposons la s\'eparation d'\'echelle en prenant un r\'eseau cristallin dont la cellule unit\'e a une longueur caract\'eristique d'ordre $m$. Les scalings des diff\'erents termes sont motiv\'es par les propri\'et\'es de changement d'\'echelle naturelles des \'energies \'electrostatiques, pour que tous les termes contribuent \'egalement \`a l'\'energie dans la limite $m\to 0$. 

Un changement d'\'echelle $\tilde\psi=m^{3/2}\psi(m\cdot)$ montre que la fonctionnelle s'exprime en variables microscopiques (i.e. du point de vue du cristal, alors que le point de vue du polaron est repr\'esent\'e par \eqref{totf1}) comme
\begin{equation}
\cE_{m}[\tilde\psi]=m^{-1}\left(\frac1{2m}\int_{\R^3}|\nabla\tilde\psi(x)|^2\,dx+\int_{\R^3}V^0_{\rm per}(x)|\tilde\psi(x)|^2\,dx+F_{\rm crys}\big[|\tilde\psi|^2\big]\right).
\label{eq:Pekar-abstract-intro-micro}
\end{equation}
Notre choix d'\'echelle revient donc \`a consid\'erer un cristal fixe et un polaron tr\`es l\'eger, de masse $m\ll 1$ (ce qui est la raison de notre choix de notation pour le rapport $m$ des \'echelles de longueur). 

Le reste de cette section explique comment retrouver la fonctionnelle de Pekar dans la limite $m\to 0$. Soulignons qu'il s'agit d'une d\'emarche tr\`es diff\'erente de la d\'erivation habituelle, \`a partir du mod\`ele de Fr\"ohlich. Ici nous regardons la transition petit polaron/grand polaron, et non la transition couplage fort/couplage faible comme dans \cite{DonVar-83,LieTho-97}.

\subsection{La matrice di\'electrique du cristal}

Le point le plus important est de comprendre comment la constante di\'electrique $\dem$, qui est un param\`etre de la fonctionnelle de Pekar, se d\'eduit du mod\`ele de cristal d\'efini \`a la Section \ref{sec:cristal}. En g\'en\'eral, cette constante est en fait une matrice, car la r\'eponse di\'electrique du cristal n'est isotrope (et donc $\dem$ scalaire) que dans le cas o\`u le r\'eseau cristallin est cubique, $\cL = \Z ^3$. Cons\'equemment nous d\'efinissons une fonctionnelle de Pekar anisotrope pour le cas o\`u $\dem$ est une matrice:
\begin{equation}\label{ptgf1}
\ptgf [\psi] :=  \frac{1}{2}\int_{\R ^{3}} |\nabla \psi| ^2 dx  +  \ptgint \big[|\psi|^2\big]
\end{equation}
avec $\ptgint$ d\'efini en variables de Fourier:
\begin{equation}\label{ptgint}
 \ptgint [\rho]: = 2\pi \int_{\R^3} \left| \hat{\rho} (k)\right| ^2 \left(\frac{1}{k^T \dem k} - \frac{1}{|k| ^2} \right)  \,dk.
\end{equation}
On peut bien s\^ur d\'efinir $\ptgf$ et $\ptgint$ en consid\'erant $W_\rho$ la solution de  l'\'equation de Poisson
\begin{equation}\label{poteff}
-{\rm div}\left( \dem \nabla \Wrho \right) = 4 \pi \rho.
\end{equation}
On a alors
\begin{equation}\label{ptgf1bis}
\ptgf [\psi] :=  \frac{1}{2}\int_{\R ^{3}} |\nabla \psi| ^2 dx  + \frac{1}{2} \int_{\R^3} |\psi|^2 \left( W_{|\psi|^2} - |\psi|^2 \star |\cdot| ^{-1}\right)
\end{equation}
et l'\'energie fondamentale correspondante est 
\begin{equation}\label{ptge}
\ptge(1) = \inf \left\lbrace \ptgf [\psi],\ \int_{\R ^3} |\psi| ^2 = 1  \right\rbrace.
\end{equation}
Comme les matrices di\'electriques satisfont toujours $\dem >1$ au sens des op\'erateurs, il est clair que l'interaction $\ptgint$ est attractive.
Pour le cas du $N$-polaron on d\'efinit de m\^eme
\begin{multline}\label{ptgfN}
\cE^{\rm P}_{\varepsilon_{\rm M}}[\Psi]=\int_{\R^{3N}}\left(\frac12\sum_{j=1}^N|\nabla_{x_j}\Psi(x_1,...,x_N)|^2+\sum_{1\leq k<\ell\leq N}\frac{|\Psi(x_1,...,x_N)|^2}{|x_k-x_\ell|}\right)dx_1\cdots dx_N\\  + \ptgint [\rhoP]
\end{multline}
et
\begin{equation}\label{ptgeN}
\ptge (N)= \inf \left\lbrace \ptgf[\Psi],\: \int_{\R ^{3N}} |\Psi| ^2 = 1, \: \Psi \mbox{ satisfaisant \eqref{eq:antisym} }  \right\rbrace.
\end{equation}
Lorsque $\dem$ est proportionnelle \`a la matrice identit\'e, on retrouve les fonctionnelles \eqref{eq:Pekar 1}-\eqref{eq:Pekar N} 

\medskip

Il existe une formule explicite pour la matrice di\'electrique d'un cristal \cite{BarRes-86,CanLew-10}, mais la d\'efinition la plus naturelle dans notre contexte passe par la d\'emarche de \cite{CanLew-10}. En r\'esum\'e, $\dem$ s'obtient en consid\'erant une excitation macroscopique de la mer de Fermi. Plus pr\'ecis\'ement, on consid\`ere un d\'efaut de charge de la forme  
\[
\nu_m(x)=m^3\,\nu(mx)                                                                                                                                                                                                                                                                                                                                                                 \]
avec $\nu$ fix\'e, c'est-\`a-dire un d\'efaut vivant \`a une \'echelle tr\`es grande devant le pas du cristal. Nous appellerons $Q_m$ une solution de ~\eqref{eq:crys def ener} associ\'ee, et $W_m$ le potentiel \'electrique correspondant, remis \`a la bonne \'echelle de longueur \footnote{Encore une fois, les normalisations des changements d'\'echelle sont motiv\'es par la r\'eponse de l'\'energie \'electrostatique \`a la dilatation des densit\'es de charge.}
\begin{equation}\label{intro:recaled potential}
W_m (x):=m^{-1}\big(\nu-\rho_{Q_m}\big)\star|\cdot|^{-1}(x/m).
\end{equation}
Canc\`es et Lewin prouvent dans \cite[Th\'eor\`eme 3]{CanLew-10} qu'il existe $\dem$ (qui peut se calculer explicitement \`a partir des donn\'ees $\munucper$ et $\cL$ caract\'erisant le cristal) telle que 
\begin{equation}\label{eq:CanLew}
W_m \wto W_{\nu}, 
\end{equation}
la solution unique de l'\'equation \eqref{poteff} avec $\rho = \nu$. Le r\'esultat \eqref{eq:CanLew} nous servira de d\'efinition pour $\dem$: il exprime le fait que le potentiel \'electrique total renvoy\'e par le cristal en r\'eponse \`a un d\'efaut de charge \'etal\'e est donn\'e par la solution de  l'\'equation de Poisson \eqref{poteff}.

Vu qu'on esp\`ere obtenir un polaron \'etal\'e \`a l'\'echelle $m ^{-1}$ en prenant la limite macroscopique pr\'esent\'ee ci-dessus, il est naturel que $\dem$ intervienne dans la fonctionnelle de Pekar. Plus pr\'ecis\'ement, on a le r\'esultat suivant qui \'eclaire le lien entre \eqref{eq:crys def ener} et \eqref{ptgint}: 

\begin{theorem}[\textbf{Comportement macroscopique de l'\'energie du cristal} \cite{LewRou-12}]\label{theo:main crys}\mbox{}\\
Soit $(\psi_m)_m$ une suite born\'ee dans $H^s(\R^3)$ pour $s>1/4$. Alors
\begin{equation}\label{eq:limit crys}
\lim_{m\to0}\Big(m^{-1}F_{\rm crys}\big[m^3|\psi_m(m\cdot)|^2\big]-F^{\rm P}_{\dem}\big[|\psi_m|^2\big]\Big)=0.
\end{equation}
\end{theorem}

Ce r\'esultat est l'\'el\'ement principal de notre d\'emarche (voir \cite{LewRou-11}). L'hypoth\`ese sur la suite $(\psi_m)_m$ est faite pour assurer une compacit\'e locale suffisante, sans laquelle on ne peut esp\'erer obtenir \eqref{eq:limit crys} (voir contre-exemple dans \cite[Section 3.3]{LewRou-11}). En effet, si la suite se concentre \`a une \'echelle trop petite, la r\'eponse du cristal sera beaucoup plus complexe. Bien s\^ur, vu le terme d'\'energie cin\'etique dans \eqref{eq:model-intro1} on aura une borne dans $H ^1$ naturelle en \'etudiant notre mod\`ele de polaron, largement sup\'erieure \`a la borne dans $H^s(\R^3)$ pour $s>1/4$ suppos\'ee ci-dessus. L'hypoth\`ese du th\'eor\`eme \ref{theo:main crys} assure cependant une compacit\'e locale dans $L ^{12/5}$, ce qui suffit pour dominer l'\'energie \'electrostatique de $|\psi_m| ^2$ et obtenir~\eqref{eq:limit crys}.

\subsection{Le mod\`ele de Pekar anisotrope et la limite macroscopique}

Avant de pr\'esenter les r\'esultats principaux de \cite{LewRou-11}, il faut remarquer que le potentiel $\FSp(./m)$ pr\'esent dans \eqref{totf1} vit \`a l'\'echelle du cristal, beaucoup plus petite que celle du polaron. Il faut donc s'attendre \`a ce que la fonction d'onde du polaron incorpore des oscillations microscopiques pour s'adapter \`a ce potentiel dont l'effet est n\'eglig\'e dans la fonctionnelle de Pekar. Il se trouve que ces oscillations peuvent \^etre prises en compte via un probl\`eme aux valeurs propres tr\`es simple: on notera $u^{\rm per}_m$ l'unique solution positive de  
\begin{equation*}
\Epere = \inf \left\lbrace\Eperf [v] \: :\: v\in H^1_{\rm per}(\Gamma), \int_{\Gamma} |v| ^2 = |\Gamma| \right\rbrace = \Eperf [\Eperm] 
\end{equation*}
avec $\Gamma$ la cellule unit\'e du r\'eseau et 
\begin{equation}\label{eq:Eperf defi}
\Eperf [v] = \int_{\Gamma} \frac{1}{2m} |\nabla v| ^2 + \FSp |v| ^2,
\end{equation}
qui donne l'\'energie d'une particule de masse $m$ tr\`es l\'eg\`ere dans le potentiel p\'eriodique~$\FSp$.

Par p\'eriodicit\'e, $\Eperm$ peut s'\'etendre \`a $\R^3$ tout entier.
Un simple argument perturbatif montre que $u^{\rm per}_m\to1$ dans $L^\ii(\R^3)$ quand $m\to0$,  et que
\begin{equation}
\Eperelim:=\lim_{m\to0} m ^{-1} \Epere = \lim_{m\to0}m^{-1}\Eperf[\Eperm] = \int_{\Gamma} V^0_{\rm per} f^{\rm per}
\label{eq:def_E_per}
\end{equation}
o\`u la fonction $f^{\rm per}\in L^\ii(\R^3)$ est l'unique solution $\cL$-p\'eriodique de 
\[
\begin{cases}
\Delta f^{\rm per}=2V^0_{\rm per} \\
\int_{\Gamma}f^{\rm per}=0.
\end{cases}
\]
Elle intervient dans le d\'eveloppement perturbatif de $u^{\rm per}_m$, voir \cite[Section 2]{LewRou-11}:
\begin{equation}\label{eq:uper expansion}
\left\|u^{\rm per}_m-1-m f^{\rm per}\right\|_{L^{\infty} (\R ^3)}\leq Cm^2.
\end{equation}
On peut en fait se d\'ebarrasser assez facilement des oscillations microscopiques par un d\'ecouplage d'\'energie. En combinant avec le Th\'eor\`eme \ref{theo:main crys} on obtient alors notre r\'esultat principal sur la d\'erivation de la fonctionnelle de Pekar:

\begin{theorem}[\textbf{La fonctionnelle de Pekar dans la limite macroscopique \cite{LewRou-11}}]\label{theo:main}\mbox{}\\
On note $\dem>1$ la matrice di\'electrique d\'efinie par \eqref{poteff} et \eqref{eq:CanLew}. Soit $N\geq1$ un entier.  
\begin{itemize}
\item\textbf{(Convergence de l'\'energie)}. On a
\begin{equation}\label{resulte1}
\boxed{\lim_{m\to0}E_m(N)=N\,\Eperelim + \ptge(N)}
\end{equation}
avec $\Eperelim$ donn\'ee par~\eqref{eq:def_E_per}, et $\ptge(N)$ est l'\'energie de Pekar donn\'ee par~\eqref{ptge} pour $N=1$ et par ~\eqref{ptgeN} pour $N\geq2$.

\bigskip

\item \textbf{(Convergence des \'etats)}. Soit $(\Psi_m)_m$ une suite de minimiseurs approch\'es pour $E_m(N)$ au sens o\`u 
$$\lim_{m\to0}\left(\cE_m[\Psi_m]-E_m(N)\right)=0.$$
On d\'efinit $\Psi^{\rm pol}_m$ par la relation
\begin{equation}\label{minimiseur pol}
\boxed{\Psi_m (x_1,...,x_N)= \prod_{j=1}^N\Eperm (x_j/m)\;\, \Psi^{\rm pol}_m (x_1,...,x_N).}
\end{equation}
Alors $(\Psi^{\rm pol}_m)_m$ est une suite minimisante pour l'\'energie de Pekar $\ptge(N)$ \eqref{ptgeN}. 

Si $N=1$ ou si $N\geq2$ et les in\'egalit\'es de liaison~\eqref{binding ptgN} sont satisfaites, il existe une suite de translations $(\tau_m)_m\subset\R^3$ et un minimiseur $\Psi^{\rm P}_{\dem}$ pour $\ptge(N)$ tels que
\begin{equation}\label{state converge}
\Psi^{\rm pol}_m (x_1-\tau_m,...,x_N-\tau_m) \to \Psi^{\rm P}_{\dem}(x_1,...,x_N) \mbox{ fortement dans } H^1 (\R ^{3N})  
\end{equation}
le long d'une sous-suite lorsque $m\to 0$.
\end{itemize}
\end{theorem}

Pour conclure, faisons deux remarques sur ce r\'esultat:
\begin{itemize}
\item Le r\'esultat \eqref{minimiseur pol} montre la s\'eparation d'\'echelles micro/macro: la fonction d'onde du polaron suit \`a l'\'echelle macroscopique un profil donn\'e par la fonctionnelle de Pekar, mais incorpore aussi des oscillations \`a l'\'echelle du cristal, d\'ecrites par $\Eperm$. On pourrait aussi bien utiliser \eqref{eq:uper expansion} pour remplacer $\Eperm$ dans l'\'enonc\'e, et au vu de cette estimation, les oscillations n'appara\^itraient pas dans une estimation $L ^2$ (ou toute autre norme n'impliquant pas de d\'eriv\'ees). Les oscillations contribuent cependant au m\^eme niveau que l'\'energie de Pekar au vu de \eqref{resulte1} et il est crucial de les extraire pour avoir le r\'esultat \eqref{state converge} dans $H ^1$.
\item En combinant avec le r\'esultat de Lewin, Th\'eor\`e\`me \ref{thm:Pekar N exist}, on voit que pour certaines valeurs de $N$ et $\dem$ (qui ne d\'epend que des donn\'ees du cristal $\munucper$ et $\cL$) et dans la limite macroscopique, les suites minimisantes pour notre mod\`ele de polaron convergent, ce qui confirme qu'il encode bien le ph\'enom\`ene de liaison, au moins pour les grands polarons. 
\end{itemize}

\bibliographystyle{siam}
\bibliography{biblio_2}

\end{document}